\documentclass{aa}  
\usepackage{eurosym,rotating, booktabs}
\usepackage{amsmath}
\usepackage{txfonts}
\usepackage{xurl}
\usepackage{amsmath}
\usepackage{amssymb}
\usepackage{longtable}
\usepackage{comment}
\usepackage{supertabular}
\usepackage{lscape}
\usepackage{multicol}

\usepackage[clockwise]{pdflscape}
\usepackage{afterpage}
\usepackage{tabularx}
\usepackage{adjustbox}

\usepackage{csquotes}
\usepackage{graphicx}
\usepackage{txfonts}
\usepackage{hyperref}
\hypersetup{
    colorlinks = true,
    linkcolor = {blue}, 
    citecolor = {blue},
    urlcolor = {blue}
}

\usepackage{threeparttable}
\usepackage{xcolor}
\usepackage{subcaption}         
\usepackage{lscape}             
\usepackage{placeins}           
\usepackage{pifont}
\newcommand{\cmark}{\ding{51}}
\newcommand{\xmark}{\ding{55}}

\begin{document}

   \title{XUE. ProDiMo models of internally and externally irradiated planet-forming disks around 0.3~--~4.0 M$_{\odot}$ stars}
   \subtitle{The IRIS project I}

   \author{Jenny Frediani\inst{1}\fnmsep\thanks{Corresponding author: jenny.frediani@astro.su.se}
        \and Konstantin V. Getman  \inst{2} 
        \and Peter Woitke  \inst{3} 
        \and Bayron Portilla-Revelo  \inst{2, 4} 
        \and Eric D. Feigelson  \inst{2} 
        \and Christian Rab  \inst{5, 6} 
        \and Arjan Bik  \inst{1} 
        \and María Claudia Ramírez-Tannus \inst{7}
        \and Rens Waters \inst{8}
        \and Thomas Henning \inst{7}
        \and Inga Kamp \inst{9}
        \and Germán Chaparro \inst{10}
        \and Sebastián Hernández \inst{10}
        \and Simón Rodríguez \inst{10}
        \and Pablo Cuartas-Restrepo \inst{10}
        \and Andrew Winter \inst{11}
        \and Thomas J. Haworth \inst{11}
        \and Thomas Preibisch \inst{5, 12}
        \and Sierk E. van Terwisga\inst{3}
        \and Veronica Roccatagliata \inst{13, 14}
        \and Alexis Brandeker \inst{1} 
        }

   \institute{Department of Astronomy, Stockholm University, AlbaNova University Centre, 106 91 Stockholm, Sweden
   \and 
   Department of Astronomy \& Astrophysics, Pennsylvania State University, 525 Davey Laboratory, University Park, PA 16802, USA
   \and 
   Space Research Institute, Austrian Academy of Sciences, Schmiedlstrasse 6, A-8042 Graz, Austria
   \and
   Center for Exoplanets and Habitable Worlds, Penn State University, 525 Davey Laboratory, 251 Pollock Road, University Park, PA 16802, USA
   \and
    University Observatory, Faculty of Physics, Ludwig-Maximilians-Universität München, Scheinerstr. 1, 81679 Munich, Germany
    \and
    Max-Planck-Institut für extraterrestrische Physik, Giessenbachstrasse 1, D-85748 Garching, Germany
    \and 
    Max-Planck-Institut für Astronomie,Königstuhl 17, D-69117 Heidelberg, Germany
    \and
    Department of Astrophysics/IMAPP, Radboud University, PO Box 9010, 6500 GL Nijmegen, The Netherlands
    \and
    Kapteyn Astronomical Institute, Rijksuniversiteit Groningen, Postbus 800, 9700AV Groningen, The Netherlands
    \and
    FACom, Instituto de Física—FCEN, Universidad de Antioquia, Calle 70 No. 52-21, Medellín 050010, Colombia
    \and
    Astronomy Unit, School of Physics and Astronomy, Queen Mary University of London, London E1 4NS, UK
    \and
    Excellence Cluster ORIGINS, Boltzmannstr.~2, 85748 Garching, Germany
    \and
    Alma Mater Studiorum, Universit\`a di Bologna, Dipartimento di Fisica e Astronomia (DIFA), Via Gobetti 93/2, 40129 Bologna, Italy
    \and 
    INAF-Osservatorio Astrofisico di Arcetri, Largo E. Fermi 5, I-50125, Firenze, Italy
   }
   


\titlerunning{XUE. ProDiMo models of internally and externally irradiated planet-forming disks}
\authorrunning{Frediani, Getman et al.}

  \abstract
    {Most stars and planets form in massive star-forming regions, where protoplanetary disks are exposed to external far-ultraviolet (FUV) radiation from nearby O- and B-type stars. It remains largely unknown, however, what is the combined effect of internal stellar irradiation and external FUV fields on the terrestrial planet-forming disk region ($\lesssim$\,10 au) across stellar masses.}
    {We model the effect of internal UV and X-ray and external FUV irradiation on mid-infrared (mid-IR) gas-phase emission, and the fraction of carbon-to-oxygen (C/O) ratio measurable in the warm disk atmosphere of T\,Tauri and Herbig Ae/Be disks.}
    {We simulate disk structures with the thermochemical code ProDiMo and process their mid-IR spectra with the FLiTs ray tracing code, convolved to the typical resolution of the James Webb Space Telescope (JWST) Mid-Infrared Instrument in Medium Resolution Spectroscopy mode ($R$\,$\sim$\,2680).}
    {(1) We build the Internal and external irRadIation of diSks (IRIS) grid, comprising four sets of models spanning stellar masses of 0.3--4.0 $M_\odot$, including stellar X-ray flares and an external FUV field of $10^4$ $G_0$ (in Habing units). (2) We systematically predict increasing flux densities with stellar mass for key atomic and molecular mid-IR diagnostic species. (3) External FUV irradiation enhances CH$_3^+$ and H$_2$ emission, while FUV-induced disk truncation retrieves internal-only irradiated inner disk chemistry. (4) Mid-IR H$_2$O, CO$_2$, and C$_2$H$_2$ line ratios imply carbon-rich compositions (C/O $\sim$1--10) of the warm emitting atmospheric layers of T\,Tauri and Herbig Ae/Be disks, primarily reflecting stellar irradiation, with little sensitivity to external FUV irradiation.}
    {The IRIS grid provides a flexible framework for interpreting JWST and potentially future Extremely Large Telescope (ELT) infrared disk observations in a large parameter space, combining central stellar and external radiation fields. Future modeling should include the impact of FUV-driven photoevaporative winds and X-ray radiative transfer, and the effects of time-dependent X-ray irradiation on inner disk chemistry.}
   

   \maketitle
   \nolinenumbers

\section{Introduction}\label{section:Introduction}
The inner regions ($\lesssim$\,10 au) of planet-forming disks, or protoplanetary disks, dictate the bulk composition of forming planets, which grow their mass upon accretion of the surrounding disk material. The gas component in these regions is typically found at temperatures of a few 100 to 1000 K, and is therefore observable in emission in the near- and mid-infrared (mid-IR) wavelength domain, limited by where in the disk the dust component becomes optically thick to the emerging thermal radiation \citep{Kamp_2004, Henning_Meeus_2011}. Various infrared surveys, particularly with the Spitzer Space Telescope (\textit{Spitzer}) \citep{Salyk_2011, Pontoppidan_2010, Pascucci_2013}, have uncovered a plethora of gas-phase molecules in inner disks, e.g., carbon monoxide (CO), water (H$_2$O), carbon dioxide (CO$_2$), acetylene (C$_2$H$_2$), and hydrogen cyanide (HCN). This points to line-rich and also H$_2$O-rich infrared spectra of the T\,Tauri solar-like disk-bearing stars (0.3\,$\lesssim$\,$M_{\star}$\,$\lesssim$\,1.5 $M_{\odot}$), of spectral type F or later \citep{Joy1945, Appenzeller1989}. These surveys also served to reveal significant chemical diversity among inner T\,Tauri disks, especially in terms of the presence of species like CO$_2$, C$_2$H$_2$, and HCN with respect to water \citep[e.g.,][]{Pontoppidan_2010}.

More recently, the increased sensitivity of JWST has allowed to extend these studies across a broader range of stellar masses and environments, uncovering an even richer chemical inventory than previously accessible \citep[e.g.,][]{Dishoeck_2023, Henning_2024, Arulanantham_2025, Arabhavi_2025, Gasman_2025, Temmink_2025_multi, Tabone_2026}. This is particularly evident for disks surrounding Very Low Mass Stars (VLMS) and brown dwarfs ($M_{\star}$ $<$\,0.3 $M_{\odot}$), exhibiting spectra surprisingly rich in hydrocarbon species \citep{Tabone_2023, Arabhavi_2024, Arabhavi_2025_water}, which were already observed at lower sensitivity by \textit{Spitzer} \citep{Pascucci_2013}. For disk-bearing stars more massive than T\,Tauri ($M_{\star}$ $\gtrsim$\,1.5 $M_{\odot}$), the so-called Herbig Ae/Be stars, and the later spectral type Intermediate Mass T\,Tauri (IMTT), precursors of the Herbigs \citep[e.g.,][]{Calvet2004, Valegard2021, Brittain2023}, both JWST and pre-JWST observations show sparse results. In particular, some exhibit very line-poor spectra \citep[e.g.,][]{Pontoppidan_2010, Fedele_2011}, while others display a volatile inventory similar to that of the lower-mass counterparts \citep{Tannus_2025, Frediani_2025, Kaufer_2026}.

While the observational picture is rapidly advancing, interpreting infrared disk spectra remains challenging. Two-dimensional (2D) thermochemical disk models are essential to take into account the radiative transfer of both the gas and the dust components, their excitation conditions, stellar host properties, and disk geometry \citep{Meijerink_2008, woitke_2009, Bruderer_2009, Bruderer_2013, Antonellini_2015, Walsh_2015, Greenwood_2019, Anderson_2021, Woitke_2018, Arabhavi_2026}, providing a computationally affordable framework for testing hypotheses about disk structure, composition, and evolution.

Although most thermochemical studies show that the host star UV and X-ray radiation is determinant for the energetics of the disk \citep[e.g.,][]{Podio_2013, Bruderer_2012, Antonellini_2016, Kama_2016, Woitke_2024_EXLupi}, the majority of stars form in clustered environments, where nearby massive OB stars expose typical disks to strong external far-ultraviolet (FUV) radiation fields of $10^{3}$--$10^{4}\,G_0$, where G$_0$ is the Habing unit \citep{Habing_1968}, driving their external UV photoevaporation and altering their evolution \citep{Winter_2020, Winter_2022, Allen2025}. This process is in fact expected to rapidly deplete disks of gas and dust and truncate them from the outside-in \citep[$\lesssim$\,1 Myr;][]{Winter_2020, Winter_2022}, shortening their lifetime in comparison to disks evolving in isolation, from roughly 3--5 Myr to 1--3 Myr \citep{Anderson_2013}. With JWST now detecting emission from volatile (gas and ice) species in disks in massive star-forming regions out to distances of 2--3 kpc \citep[e.g.,][]{Berne_2023, Tannus_2025, Guarcello_2025, Potapov_2025, Ballering_2025}, external FUV irradiation has become an essential ingredient in the interpretation of infrared disk spectra in such regions.

Several model investigations of externally irradiated disks have been done of the outer planet-forming regions ($>$\,10 au), motivated by far-infrared and sub-millimeter observations \citep{Walsh_2013, Boyden_2023, Gross_2025, Keyte_2025, Keyte_2026}, and of the inner regions as well \citep{Walsh_2013, Portilla-Revelo_2025, Calahan_2025, Hernandez_Arboleda_2026}. However, the investigations of the latter have all been focusing on the T\,Tauri regime, and cover an inhomogeneous parameter space of disk thermochemical structure and also of the analyzed volatile species, especially atomic and molecular gas, observable in these inner regions.
One of the key missing pieces in our understanding of planet formation and planet compositions therefore is a systematic exploration of how internal stellar irradiation and external FUV fields jointly shape the observable mid-IR emission and the chemistry in the inner regions of planet-forming disks.

We address this problem in the IRIS (Internal and External Radiation of Planet-forming Disks) project. In this first paper, we introduce the IRIS grid of 2D thermochemical ProDiMo models spanning stellar masses from 0.3 to 4.0 $M_\odot$, while varying both the internal stellar irradiation in terms of UV and X-ray input and the external FUV radiation field, following observed scaling relations in the known population of protoplanetary disks, as visually summarized in Table~\ref{table:summary_grid}. In this parameter space, we focus on the gas-phase content in the innermost $\lesssim$10 au disk region, in particular, the mid-IR rovibrational emission coming from key atomic (H\,\textsc{i}) and molecular (H$_2$, CO, H$_2$O, OH, CO$_2$, C$_2$H$_2$, HCN, CH$_3^+$) species, as also observable with the JWST Mid-InfraRed (MIRI) instrument. Setting the stage for the near-future Extremely Large Telescope (ELT), this model grid provides a framework for interpreting present and future JWST observations of T\,Tauri and Herbig Ae/Be disks forming in both isolated and clustered environments, with a focus on their mid-IR spectra, diagnostics species, and the carbon-to-oxygen (C/O) ratio inferrable from infrared observations. The latter directly relates to the chemical composition during planet formation, as the relative volatility of carbon- and oxygen-bearing species across disk snow lines sets the C/O ratio inherited by forming planets \citep{Oberg_2011}. Recent JWST observations have revealed that the inner-disk C/O ratio varies with stellar properties, with carbon-rich spectra found preferentially around very low-mass stars \citep{Tabone_2023, Arabhavi_2024}, a trend that thermochemical models have attributed to internal photoevaporation, which may limit the inward transport of oxygen-rich material \citep{Lienert_2024, Lienert_2025}. Other observational studies link this chemical diversity not only to stellar but also outer-disk properties, offering a demographic view of how volatile transport shapes planet-forming regions \citep[e.g.,][]{Pontoppidan_2024, Arulanantham_2025}.

As we lack observational constraints for VLMS in high-mass star-forming regions, and their modeling would require additional careful consideration of the assumptions about disk structure, we exclude from our modeling disks surrounding lower-mass stars than T\,Tauri. In a separate paper, we benchmark the IRIS grid against high quality JWST/MIRI spectra of 33 known T\,Tauri and Herbig Ae/Be disks located both in isolated and externally irradiated environments (Frediani et al., in prep).

The paper is organized as follows. The modeling setup and the sets of disk models are explained in Sect. \ref{section:Model setup}. Sect. \ref{section:Methodology} illustrates the methods used to analyze and interpret the IRIS models. The key results are presented and discussed in Sects. \ref{section:Results} and \ref{section: Discussion}, and the conclusions of our modelling work are summarized in Sect. \ref{section:Conclusions}.

\begin{table}[htp!]
    \caption{Visualization of the irradiation properties of the model sets included in the IRIS grid.}
    \centering
    \resizebox{\columnwidth}{!}{
    \begin{tabular}{lcccl}
    \hline
    \hline
    \noalign{\vskip 1mm}
    Model set & UV & X-ray & FUV & Disk size\,$^{b}$ \\
              & (stellar) & (stellar) & (external) & \\
    \noalign{\vskip 1mm}
    \hline
    \noalign{\vskip 0.8mm}
    \texttt{Baseline} & \cmark & \cmark & \xmark & 100 au \\
    \texttt{X-ray Flare} & \cmark & \cmark \cmark \,$^{a}$ & \xmark & 100 au \\
    \texttt{Ext. UV} & \cmark & \cmark & \cmark & 100 au \\
    \texttt{Trunc. Ext. UV} & \cmark & \cmark & \cmark & 10 au \\
    \hline
    \end{tabular}}
    \begin{tablenotes}
    \footnotesize
    \item \textbf{Notes.} $^{a}$\,A stellar X-ray superflare is included. $^{b}$\,It refers to the radial extension of the gas component before exponential cutoff of the surface density profile.
    \end{tablenotes}
    \label{table:summary_grid}
\end{table}
\section{Modeling setup}\label{section:Model setup}
In order to simulate the terrestrial planet-forming region of disks as a function of stellar mass and radiation input, we use the Protoplanetary Disk Model code \citep[ProDiMo\footnote{Master branch revision \texttt{f3963b27}, \url{https://prodimo.iwf.oeaw.ac.at/}},][]{woitke_2009, kamp_2010, thi_2011, rab_2018, Woitke_2024_EXLupi}. Section \ref{section: fiducial models} describes in detail the fundamental setup of the physical and chemical properties of our grid of ProDiMo models,  while Sect. \ref{subsection: Baseline} to \ref{subsection: Ext. UV with trunc.} focus on the different disk irradiation recipes included in the grid.

The workflow of the code can be summarized as follows. It first sets up the physical structure of the disk, including the emission from the pre-main-sequence star, by initializing, for example, the gas and dust density distributions and the stellar spectrum. Then, according to the chosen dust species and size distributions, the dust opacities are computed in order to solve the continuum radiative transfer, and determine the final dust temperature structure. The code proceeds by calculating the gas temperature structure and the abundance of different atomic and molecular species coupled with the dust. This is done by computing the disk chemistry on the basis of a chemical rate network of reactions, and solving the radiative balance (heating and cooling) for those species. 

We adopt the large chemical network and elemental abundances from \citet{kamp_2017}, while the creation and destruction processes with respective photo-reaction rate coefficients are mainly taken from the UMIST 2022 database of \citet{Millar_2024}, combined with updated escape probability and molecular self-shielding factors \citep{Woitke_2024_EXLupi}. In particular, the photo cross-sections to calculate photodissociation and photoionization rates in disk surface layers are taken from \citet{Heays_2017} and \citet{Hrodmarsson_2023}. 

For the atomic and molecular species analyzed in our grid we use Local Thermodynamic Equilibrium (LTE) spectroscopic data (CH$_3^+$, C$_2$H$_2$), or, when available, non-LTE dedicated model data (H\,\textsc{i}, H$_2$, CO, H$_2$O, OH, CO$_2$, HCN). The main source databases are HITRAN \citep{gordon_2022} and LAMDA \citep{vandertak_2020}. The specific line list reference(s) for each species included in the IRIS grid can be found in Sect. \ref{section: selection of disk observables}. All level populations (from LTE and non-LTE data) are computed in ProDiMo using the escape probability formalism \citep{Woitke_2024_EXLupi}, which measures the fraction of emitted photons that ultimately escape an emitting region without being re-absorbed. We then postprocess the ProDiMo outputs with the Fast Line Tracer \citep[FLiTs\footnote{\url{https://github.com/michielmin/FLiTs}},][]{Woitke_2018} code to do a 3D photon ray tracing of the computed level populations, solving the radiative-transfer problem along many lines of sight in the disk. In particular, FLiTs produces line emission spectra for a given inclination angle, including the effect of dust continuum and line blending, i.e., physically overlapping spectral lines of different species. 

Combining ProDiMo with FLiTs, we ultimately obtain synthetic mid-IR spectra between 4.9 and 28 $\mu$m and we convolve them to a resolving power of $R$ = 2680, which are respectively representative of the wavelength range and the resolution covered by JWST/MIRI at 15 $\mu$m in Medium Resolution Spectroscopy (MRS) mode \citep{argyriou_2023}. Our ProDiMo models are therefore tailored to JWST spectroscopy, as demonstrated in a separate IRIS paper (Frediani et al., in prep.), but can readily be adapted to simulate observations with other facilities, such as the ELT.

\subsection{The IRIS grid}\label{section: fiducial models}
The IRIS grid comprises 45 ProDiMo simulations, grouped into four model sets spanning central pre-main-sequence stellar masses from 0.3--4.0 $M_{\odot}$ (see Table~\ref{table:summary_grid}). The fundamental disk and host star parameters set in the grid are described separately in Sect. \ref{disk parameters} and \ref{star parameters}.

\begin{table}[htp!]
    \caption{Assumed disk ProDiMo parameters fixed for all sets of models in the IRIS grid.} \label{table: structural parameters}
    \centering
    {%
    \begin{tabular}{ll}
    \hline
    \hline
    \noalign{\vskip 1mm}
    Parameter & Value\\
    \noalign{\vskip 1mm}
    \hline
    \noalign{\vskip 1mm}
    Grid resolution (NXX $\times$ NZZ) & 150 $\times$ 100\\
    \hline
    \noalign{\vskip 0.8mm}
    Gas mass ($M_{\rm disk}$)$^{\,a,b}$ & 1\% of $M_{\star}$\\
    Dust-to-gas mass ratio ($d/g$)$^{\,b}$ & 10$^{-2}$ \\
    Tapering-off radius ($R_{\rm tap}$)$^{\,b}$ & 100 au\\
    Reference radius ($R_{\rm ref}$) & 1 au\\
    Flaring index ($\beta$)$^{\,c}$ & 1.15\\
    Scale height at $R_{\rm ref}$ ($H_{0}$)$^{\,c}$ & 0.051\\
    Line-of-sight inclination angle & 60$^{\circ}$\\
    \noalign{\vskip 0.8mm}
    \hline
    \noalign{\vskip 0.8mm}
    C/O ratio & 0.457\\
    O/H ratio & 3.02 $\times$ 10$^{-4}$ \\
    C/H ratio & 1.38 $\times$ 10$^{-4}$\\
    Column density power index ($\epsilon$) & 1.0 \\
    Cosmic ray H$_2$ ionisation rate & 10$^{-17}$ s$^{-1}$\\
    \noalign{\vskip 0.8mm}
    \hline
    \noalign{\vskip 0.8mm}
    Min. dust grain size ($a_{\rm min}$) & 0.05 $\mu$m\\
    Max. dust grain size ($a_{\rm max}$) & 3000 $\mu$m\\
    Grain porosity & 25\% \\
    Dust composition & Mg$_{0.7}$Fe$_{0.3}$SiO$_{3}$ (60\%) \\
    (volume fraction) & amorph. carbon (15\%)\\
    Dust distribution index ($a_{\rm pow}$) & 3.5\\
    PAHs abundance ($f_{\rm PAH}$) & 10$^{-2}$ \\
    Turbulent mixing parameter ($\alpha_{\rm settle}$) & 10$^{-3}$\\
    Dust settling prescription & \citet{Riols_2018}\\
    \noalign{\vskip 1mm}
    \hline
    \end{tabular}
    }
    \begin{tablenotes}
    \footnotesize
    \item \textbf{Notes.} $^{a}$\,The disk mass therefore linearly increases with increasing stellar mass model. $^{b}$\,$M_{\rm disk}$, $d/g$ ratio, and $R_{\rm tap}$ are then modified to simulate disk truncation under external FUV photoevaporation; see Sect. \ref{subsection: Ext. UV with trunc.}. $^{c}$\,These parameters are computed as a function of $R_{\rm ref}$; see text.
    \end{tablenotes}
\end{table}
\subsubsection{Disk structural parameters}\label{disk parameters}
Our aim is to vary the radiation input on the disk in terms of UV and X-rays coming from the star, and of FUV photons coming from the external environment (nearby OB stars) across the different sets of models. To isolate the effects of irradiation on the inner disk structure and its emerging mid-IR spectrum, we firstly assume in all sets of models the properties reported in Table~\ref{table: structural parameters}, which are mainly related to the geometry, elemental abundances, and dust properties of the simulated disk structures. 

The values of disk mass ($M_{\rm disk}$), dust-to-gas mass ratio ($d/g$), and the tapering-off radius ($R_{\rm tap}$), which defines the exponential cutoff of the disk surface density, are motivated by the trends observed in the known disk population in nearby regions \citep[e.g.,][]{Andrews_2013, Manara_2023}, and by standardized ProDiMo models \citep{Woitke_2016}. The values of these parameters are fixed in the whole grid except for one model set of the IRIS grid, where we modify them to mimic disk truncation and outer gas depletion in presence of FUV external photoevaporation (Sect. \ref{subsection: Ext. UV with trunc.}). Knowing the disk mass and the tapering radius, the outer disk radius ($R_{\rm out}$) is then computed in ProDiMo such that the column density of neutral hydrogen equals 10$^{20}$ cm$^{-2}$ at this radius. Thus, $R_{\rm out}$ in the IRIS grid is approximately 520--600 au for stellar masses of 0.3–4.0 \(M_{\odot}\), whereas it is fixed at 300 au in the truncated model set (Sect. \ref{subsection: Ext. UV with trunc.}). Owing to the tapered surface-density profile, only a negligible fraction of the disk mass resides beyond the tapering-off radius. As our analysis focuses on the inner disk regions, these relatively large outer radii do not affect the results.

The scale height is parametrically set as $H(r) = H_0 (r/R_{\rm ref})^{\,\beta}$, where $\beta$ is the flaring index. We choose a reference radius of $R_{\rm ref}\!=\!1\,$au to directly set the disk scale height at radii typically probed by JWST observations. The index $\beta$ and the reference scale height $H_0$ at the reference radius are then set accordingly at 1 au (Table~\ref{table: fixed parameters}).

The inner disk radius ($R_{\rm in}$) scales with the stellar mass and is therefore computed for each stellar mass model. The resulting values are summarized together with the host star parameters in Table~\ref{table: fixed parameters}. $R_{\rm in}$ is identified both for the dust and the gas component with the silicate dust sublimation radius, i.e., the radial distance at which the equilibrium temperature of silicate grains reaches 1500 K, inside which silicates sublimate \citep{Dullemond_2001, Woitke_2024_CAI}. This definition places the inner disk rim sufficiently far from the star to avoid unrealistically high midplane temperatures that could affect the heating and cooling processes in the inner disk.

We note that in our models the gas inner radius is tied to the dust inner radius: inside the dust sublimation radius, the steady-state axisymmetric disk model in hydrostatic equilibrium and Keplerian rotation is no longer physically applicable. The sublimated dust removes the main opacity source and cooling mechanism of the disk, and the region is dominated by magnetospheric accretion and free-falling gas onto the star, which is a fundamentally different physical regime than what ProDiMo can treat.

We also use fixed elemental abundances of carbon-to-hydrogen (C/H) and oxygen-to-hydrogen (O/H) corresponding to a solar bulk C/O = 0.457, together with a column-density power-law index of $\epsilon$ = 1 \citep{woitke_2009}, describing how column densities change with radius. The cosmic ray H$_2$ ionisation rate is set to the commonly adopted value of $10^{17}$\,s$^{-1}$ used in disk chemical models \citep{woitke_2009,Padovani_2018}.

For the dust properties we rely on the prescriptions from the DIANA project \citep{Woitke_2016}, and the \citet{Riols_2018} prescription for dust settling, which is turbulence-based and calibrated against magneto-hydrodynamic simulations. A fraction of the disk’s carbonaceous material is present as polycyclic aromatic hydrocarbons (PAHs). Their abundance is set in ProDiMo as a scaling factor relative to the assumed standard interstellar medium PAH abundance \citep{Woitke_2016}, which is valued 3 $\times$ 10$^{-7}$ relative to hydrogen nuclei. This is the typical value observed in protoplanetary disks around low- and intermediate-mass stars \citep{Geers_2006, Tielens_2008}. For the IRIS grid, we choose to fix $f_{\rm PAH}$ = 10$^{-2}$ (in Table~\ref{table: structural parameters}), motivated by their general low detection rate found in T\,Tauri disks, and from thermochemical models of irradiated cases \citep[e.g.,][]{Geers_2006, Seok_2017, Hernandez_Arboleda_2026}.

\begin{table*}[htp!]
    \caption{Assumed host star ProDiMo parameters across the IRIS grid.} \label{table: fixed parameters}
    \centering
    {%
    \begin{tabular}{ccccccccccc}
    \hline
    \hline
    \noalign{\vskip 1mm}
    $M_{\star}$ &  $T_{\rm eff}^{~(1)}$ & $R_{\star}$ & $R_{\rm in}$ & $L_{\star}^{\,(1)}$ & $L_{\rm acc}^{\,(3)}$ & $L_{\rm UV/\star}^{\,(2)}$ & $\dot{M}_{\rm acc}^{\,(4),\,a}$ & $L_{\rm X}^{(5)}$ & $L_{\rm X,\,flare}^{(5)}$ & $T_{\rm X}^{~(6),\,b}$\\
    \noalign{\vskip 0.8mm}
    $[M_{\odot}]$ & [K] & [$R_{\odot}$] & [au] & $[L_{\odot}]$ & $[L_{\odot}]$ & [-] & $[M_{\odot}$/yr] & [erg/s] &  [erg/s]  & [MK]\\
    \noalign{\vskip 1mm}
    \hline
    \noalign{\vskip 1mm}
    0.3 & 3437 & 2.05 & 0.05 & 0.53 & 0.04 & 0.013 & 2.20 $\times 10^{-8}$ & 3.63 $\times 10^{29}$ & 1.25 $\times 10^{31}$ & 24.06\\
    0.5 & 3828 & 2.16 & 0.07 & 0.90 & 0.10 & 0.018 & 3.27 $\times 10^{-8}$ & 8.58 $\times 10^{29}$ & 1.29 $\times 10^{31}$ & 25.30\\
    0.7 & 4178 & 2.37 & 0.09 & 1.54 & 0.25 & 0.026 & 6.09 $\times 10^{-8}$ & 1.51 $\times 10^{30}$ & 1.34 $\times 10^{31}$ & 26.16\\
    0.9 & 4485 & 2.57 & 0.13 & 2.40 & 0.52 & 0.035 & 1.05 $\times 10^{-7}$ & 2.32 $\times 10^{30}$ & 1.41 $\times 10^{31}$ & 26.82\\
    1.5 & 4861 & 3.24 & 0.22 & 5.27 & 1.96 & 0.060 & 2.90 $\times 10^{-7}$ & 5.48 $\times 10^{30}$ & 1.68 $\times 10^{31}$ & 28.21 \\
    2.0 & 5047 & 3.75 & 0.28 & 8.24 & 4.16 & 0.081 & 5.30 $\times 10^{-7}$ & 8.88 $\times 10^{30}$ & 1.98 $\times 10^{31}$ & 29.03\\
    3.0 & 5340 & 4.90 & 0.45 & 17.62 & 14.90 & 0.136 & 1.64 $\times 10^{-6}$ & 1.76 $\times 10^{31}$ & 2.73 $\times 10^{31}$ & 30.22\\
    3.5 & 5564 & 6.18 & 0.67 & 32.96 & 42.67 & 0.208 & 5.01 $\times 10^{-6}$ & 2.27 $\times 10^{31}$ & 3.18 $\times 10^{31}$ & 30.68\\
    4.0 & 6205 & 8.36 & 1.31 & 93.32 & 245.15 & 0.423 & 5.73 $\times 10^{-6}$ & 2.85 $\times 10^{31}$ & 3.68 $\times 10^{31}$ & 31.09\\
    \noalign{\vskip 1mm}
    \hline
    \end{tabular}
    }
    \begin{tablenotes}
    \footnotesize
    \item \textbf{Notes.} $^{a}$\,Refined in ProDiMo as a function of the inner disk radius; refer to Sect. \ref{section: fiducial models}. $^{b}$\,These values are then increased and used together with the $L_{\rm X,\,flare}$ column to simulate the inclusion of a X-ray superflare of the star; see Sect. \ref{subsection: Xray Flare}.
    \item \textbf{References.} (1) 1 Myr PARSEC isochrone \citep{Nguyen_2022}; (2) $L_{\rm \,acc}$--$L_{\rm bol}$ relation from \citet{Manara_2023}; (3) $L_{\rm FUV,\,acc}$--$L_{\rm acc}$ relation using a blackbody approximation for accretion hotspots of $T_{\rm acc}$ = 10000 K \citep{Calvet_1998, Herczeg_2014}; (4) $\dot{M}_{\rm acc}$--$M_{\star}$ relation from \citet{Manara_2023}; (5) $L_{\rm X}$--$M_{\star}$ relation from \citet{Telleschi_2007}; (6) $T_{\rm X}$--$M_{\star}$ relation from \citet{Getman_2005}.
    \end{tablenotes}
\end{table*}
\subsubsection{Host star parameters}\label{star parameters}
The properties of the central host star determine the irradiation that continuously impinges on the surrounding disk. To constrain these stellar properties, summarized in Table~\ref{table: fixed parameters}, we adopt empirical trends observed in the population of known pre-main-sequence stars with ages of approximately 1 Myr to fix the bolometric luminosity ($L_{\star}$), intended as photospheric luminosity of the host star, stellar effective temperature ($T_{\rm eff}$), accretion luminosity ($L_{\rm acc}$), the ratio between UV and total bolometric luminosity ($L_{\rm UV/\star}$), mass accretion rate ($\dot{M}_{\rm acc}$), stellar X-ray luminosity ($L_{\rm X}$), and of the X-ray emission temperature \citep[$T_{\rm X}$;][]{Getman_2005, Telleschi_2007, Nguyen_2022, Manara_2023}. The latter controls the shape or hardness of the stellar X-ray spectrum. Assuming that stellar X-rays originate from an optically thin thermal plasma, $T_{\rm X}$ is fixed by default in ProDiMo to 4 MK to represent a quiescent coronal temperature. 

We adopt a protostellar system age of 1 Myr primarily to investigate the gas-rich phase of disk evolution, during which conditions are particularly suitable to planet formation. This age is especially relevant because external FUV photoevaporation can deplete the disk gas on comparable or shorter timescales, thereby competing with planet formation for the available gas reservoir. Since our model setup is also representative of the typical evolutionary stage of the comparison sample of T\,Tauri and Herbig Ae/Be disks discussed in a separate paper (Frediani et al., in prep.), we expect our results to remain unvaried over an age range of $\sim$0.7--3 Myr.

For our set of stellar masses of interest, the corresponding $T_{\rm eff}$ and $L_{\star}$ values are therefore estimated assuming a stellar age of 1 Myr and using the PARSEC evolutionary models \citep{Nguyen_2022}. The bolometric luminosities and the effective temperatures are then used to set the stellar radius $R_{\star}$ using the Stefan–Boltzmann law ($L_{\star}$ = 4$\pi$ $R_{\star}^2$ $\sigma\,T_{\rm eff}^4$). 

The accretion luminosities ($L_{\rm acc}$) and mass accretion rates ($\dot{M}_{\rm acc}$) are estimated from the empirical $L_{\rm acc}$--$M_{\star}$ and $\dot{M}_{\rm acc}$--$L_{\rm acc}$ relations derived from the stellar and disk properties compiled by \citet{Manara_2023} for the nearby T\,Tauri disk population, and extrapolated into the Herbig Ae/Be regime. For these stellar masses, our estimated $\dot{M}_{\rm acc}$ are comparable with the new compilation for the population of Herbig Ae/Be disks within 1 kpc \citep{Stapper_2026}. We refine the values of $\dot{M}_{\rm acc}$ as a function of the computed $R_{\rm in}$ across the grid following the relation:
\begin{equation}
    L_{\rm acc} = \frac{GM_{\star} \dot{M}_{\rm acc}}{2 R_{\star}} \left( 1-\frac{R_{\star}}{R_{\rm in}} \right)
\end{equation}
where $G$ is the gravitational constant, while the other quantities are known for each model (see Table~\ref{table: fixed parameters}). We note that the scatter in the observed mass accretion rates as a function of stellar masses is particularly significant ($>$\,1 dex), and it is similarly observed for disk masses. As a sanity check, we run preliminary ProDiMo models with the \texttt{Baseline} setup and increased (decreased) $\dot{M}_{\rm acc}$ and $M_{\rm disk}$ by a factor $\times$\,100 ($\times$\,0.01), and found negligible differences in the output mid-IR spectra, flux densities, and molecular abundances with respect to the values fixed in the IRIS grid for these parameters.

The stellar UV luminosities (integrated over the 90--250 nm band) associated with accretion processes are estimated from $L_{\rm acc}$ by assuming that the accretion hotspot of the pre-main sequence star emits as a blackbody with a temperature of $\approx10{,}000$ K \citep{Calvet_1998, Herczeg_2008, Tofflemire_2017}. The $L_{\rm UV}/L_{\rm bol}$ column in Table~\ref{table: fixed parameters} hence reports the ratio between the computed bolometric and UV luminosities. X-ray luminosities ($L_{\rm X}$) and average coronal plasma temperatures ($T_{\rm X}$) are estimated from the empirical $L_{\rm X}$--$M_{\star}$ and $T_{\rm X}$--$M_{\star}$ relations derived from observations of pre-main-sequence stars in the Orion Nebula Cluster and the Taurus Molecular Cloud \citep{Getman_2005, Preibisch_2005, Telleschi_2007}, albeit a large known scatter \citep{Preibisch_2005}. We also include the EUV energy range (12–91 nm) for the X-ray chemistry treatment in our models. Unless X-ray photon scattering is explicitly included in ProDiMo \citep{rab_2018}, the X-ray chemistry module runs by default with the X-ray ionization rates and the resulting chemical reactions based on the treatment originally implemented by \citet{Aresu_2011}.

The IRIS grid simulates gas-rich disks around $\sim$1 Myr-old \textit{FGKM}-type T\,Tauri and intermediate-mass T\,Tauri (IMTT) pre-main-sequence stars, with a focus on the terrestrial planet-forming regions accessible to JWST observations. The disks are assumed to be smooth, without substructures, and initially extended to $\sim$100 au before the outer cutoff, with standard prescriptions for turbulence, dust composition, and settling. Internal and externally driven photoevaporative winds at the disk surface are neglected. Within this framework, we explore the effects of internal and external UV and X-ray irradiation through four model sets: a reference set including only stellar UV and X-ray irradiation (Sect. \ref{subsection: Baseline}); a set incorporating a stellar X-ray superflare (Sect. \ref{subsection: Xray Flare}); and two sets exposed to an external FUV field of $10^{4}\,G_0$, applied to an extended and a truncated disk, respectively (Sects. \ref{subsection: Ext. UV} and \ref{subsection: Ext. UV with trunc.}). Modeling caveats and possible future improvements to the IRIS grid are discussed in Sect. \ref{discussion: model limitations}.

\begin{figure*}[!htbp]
    \centering
\includegraphics[width=\linewidth]{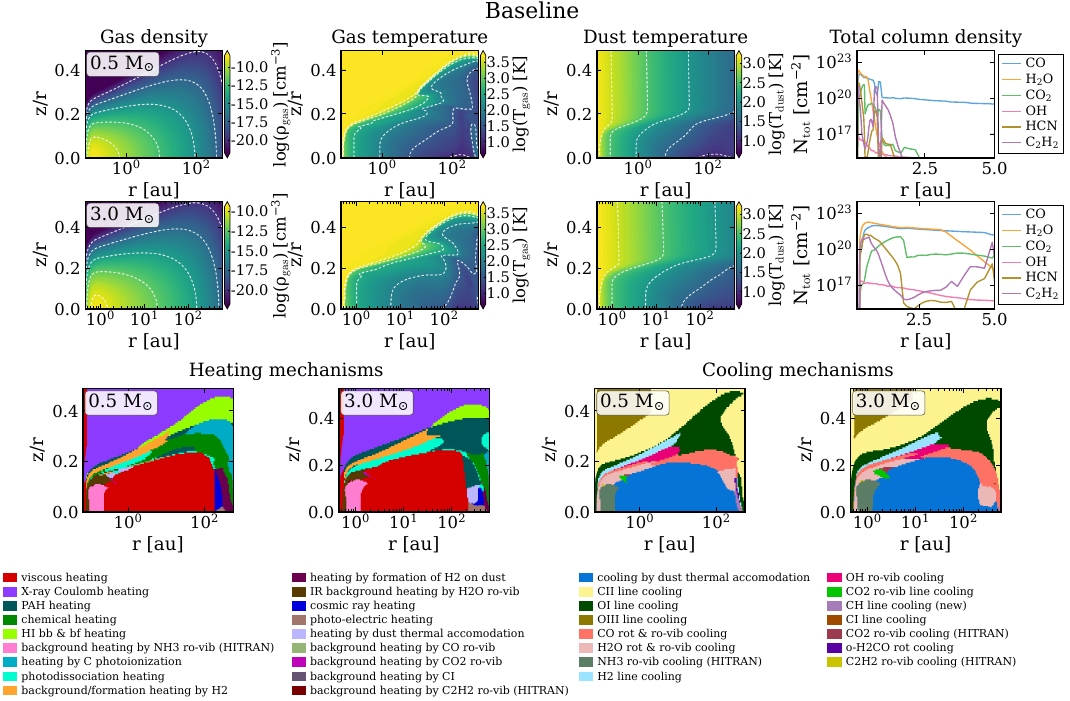}
    \caption{Overview of the physical and chemical structure of the \texttt{Baseline} set of IRIS grid models for two representative stellar masses ($M_{\star}$ = 0.5, 3.0 $M_{\odot}$). The \textit{x}-axis is the radial distance from the star in units of au, and the \textit{y}-axis the disk height above the midplane (dimensionless). The top two rows show from left to right the 2D gas density structure ($\rho_{\rm gas}$), gas temperature structure in equilibrium with the dust ($T_{\rm gas}$), and dust temperature structure ($T_{\rm dust}$) versus logarithmic radius, and the total vertical column density as a function of the radial distance in linear scale from the star ($N_{\rm tot}$) for some of the species analyzed in this work (see legend and Table~\ref{table: lines and rms}). Panels in the bottom row summarize from left to right the dominant heating and cooling mechanisms (see legends) regulating the 2D structure. Mechanisms are sorted from the most dominant to the least dominant one.}
    \label{figure: baseline grid properties}
\end{figure*}
\subsection{Model set: Baseline}\label{subsection: Baseline}
The first set of IRIS models (hereafter, \texttt{Baseline}) provides the reference for comparison with all subsequent model sets. It includes only the stellar UV and X-ray irradiation parameters listed in Table~\ref{table: fixed parameters}, namely $L_{\star}$, $L_{\rm UV}/L_{\star}$, $L_{\rm acc}$, $L_{\rm X}$, and $T_{\rm X}$, with no stellar X-ray flaring or external FUV irradiation.

This model set represents the case of smooth and extended protoplanetary disks irradiated with UV and X-rays from an increasingly massive and bright \textit{FGKM} pre-main sequence star. Fig.~\ref{figure: baseline grid properties} illustrates the emerging 2D disk structure of such physical setup for two representative models in the grid characterized by a host star with mass $M_{\star}$ = 0.5 $M_{\odot}$ and 3.0 $M_{\odot}$ respectively. First two rows show from left to right the 2D gas density profile ($\rho_{\rm gas}$), gas ($T_{\rm gas}$) and dust ($T_{\rm dust}$) 2D temperature profiles, and the total column density ($N_{\rm tot}$) of some of the gas-phase species analyzed in this work as a function of the radial distance from the star. We emphasize that the plotted column density is integrated as a function of disk radius over its entire vertical extent, and therefore quantifies the entire column of gas present in the disk, also below the $\tau_{\rm dust}(\lambda) = 1$ optically thick dust surface, which cannot be probed observationally. The bottom row in Fig. \ref{figure: baseline grid properties} shows the main active radiative heating and cooling mechanisms in the disk structure included with ProDiMo.

The density of the gas roughly follows a power-law relation as a function of radius, particularly close to the central midplane. Similarly, as a result of the radiative transfer calculation with heating and cooling physics, the structure of the gas and dust temperatures are also well-described by power-laws. In particular, the innermost ($r$ $<$ 1 au, $z/r$ $<$ 0.2) layers are the densest and warmest, while the outermost ones, especially towards the disk midplane, have lower density and are colder ($r$ $\geq$ 50 au, $z/r$ $<$ 0.1 in figure). In the low-mass models (for $M_{\star}$ $<$ 2.0 $M_{\odot}$) the vertical gas-phase abundance of the main molecular carriers peaks within $\approx$2 au. CO remains instead the most abundant molecule, and its emission extends roughly over the entire disk structure (see $N_{\rm tot}$ panels in Fig. \ref{figure: baseline grid properties}). 

Moving to higher stellar host masses ($M_{\star}$\,$\gtrsim$\,2.0 $M_{\odot}$), the disk becomes warm enough to keep the same molecular species in the gas at larger radii, with an abundance cutoff at $\approx$5--6 au. In addition to that, the gas-phase abundance of C$_2$H$_2$ and HCN peak in two distinct regions, one located roughly around 1 au, and the second one arising at progressively larger radius (around 5 au for $M_{\star}$ = 3.0 $M_{\odot}$ and 8 au for higher stellar mass models). Meanwhile, the CO$_2$ gas-phase abundance starts to peak at around 2 au, remaining constant until 4--6 au before cutting off. 

In terms of specific heating and cooling mechanisms (see bottom panels in Fig. \ref{figure: baseline grid properties}), the disk structure is largely dominated as expected by dust viscous heating and its cooling by thermal accommodation, while the inner regions specifically are regulated by water (H$_2$O), ammonia (NH$_3$), and molecular hydrogen (H$_2$) heating and cooling through roto-vibrational line emission or absorption. This behaviour reflects the combination of their rich rovibrational spectra, relatively large permanent dipole moments, and efficient collisional coupling at the high gas densities of the inner disk.

As the stellar host mass increases, the stellar luminosity and its UV flux become stronger, shifting the dominant heating and cooling processes. CO rovibrational heating and cooling become increasingly important because large volumes of the disk start to reach sufficiently high temperatures to populate higher rotational and vibrational CO levels. At the same time, the contribution of PAH photoelectric heating and photodissociation also increase with stellar mass, owing to the stronger stellar FUV radiation field. In particular, FUV photons eject energetic electrons from PAH grains, which can then transfer their kinetic energy to the gas via collisions. The increased fluxes of dissociating FUV photons also penetrate a larger fraction of the disk molecular layer, leading to more efficient destruction of molecules such as H$_2$O, NH$_3$, HCN, and C$_2$H$_2$.

\subsection{Model set: X-ray Flare}\label{subsection: Xray Flare}
In the model set hereafter designated \texttt{X-ray\,Flare}, we add an X-ray luminosity of $7 \times 10^{30}$ erg s$^{-1}$ to the baseline stellar X-ray emission of each model ($L_{\rm X,flare}$ in Table~\ref{table: fixed parameters}). This additional component corresponds to the average luminosity of pre-main-sequence X-ray superflares with total X-ray energies of $\log(E_{\rm X}/\mathrm{erg}) \sim 34$--$35$ \citep{Getman_2008, Getman_2021a, Getman_2021b}. Such pre-main sequence star X-ray super-flares are expected to occur approximately weekly, making them much more frequent than the more energetic X-ray mega-flares ($\log(E_{\rm X}$) \,$\sim$\,$36$--$38$ erg), which occur on timescales of several months to years. Likewise, in this model set we increase the temperature of the X-ray-emitting coronal plasma ($T_{\rm X}$) to 60 MK to mimic the persistently hotter plasma characteristic of pre-main sequence star X-ray super-flares. As a first-order approximation of a stellar X-ray super-flare, in our steady-state models we introduce a constant enhancement of the stellar X-ray luminosity and a harder X-ray spectrum.

The output disk structure is exemplified in Fig. \ref{figure: xrayFlare grid properties}, and is overall similar to the baseline case. The increased stellar X-ray luminosity still has some effect on the low mass models ($M_{\star} \lesssim$\,1.5 $M_{\odot}$), and for the high mass models ($M_{\star} >$\,1.5 $M_{\odot}$) separately. In the former subset of low-mass models, the gas-phase column density of C$_2$H$_2$ and HCN appear to be relatively suppressed ($N_{\rm tot} <$ 10$^{18}$ cm$^{-2}$) in comparison to the baseline models (see right-most plot in first and second row of Fig. \ref{figure: xrayFlare grid properties}). This is likely due to the fact that the X-ray photodestruction rates
of C$_2$H$_2$ and HCN are larger than their gas-phase formation rates. At the same time, a higher X-ray luminosity contributes to make H$_2$O and H$_2$ more important contributors, i.e., over larger radii, in the heating and cooling balance of the disk, thanks to the presence of extra X-ray-produced free electrons, and higher X-ray heated gas temperatures (see bottom panels in Fig. \ref{figure: xrayFlare grid properties}). 

For models with $M_{\star} >$\,1.5 $M_{\odot}$, the primary effect is instead the formation of a more extended gas-phase reservoir of C$_2$H$_2$ and HCN at larger radii (see the bottom panel of Fig. \ref{figure: xrayFlare grid properties}, showing $N_{\rm tot}$). This is likely because the enhanced X-ray photodestruction rates are compensated by stellar UV radiation, which promotes the gas-phase formation of C$_2$H$_2$ and HCN. Overall, though, the increased X-ray properties mimicking the presence of a stellar X-ray super-flare do not cause dramatic changes to the inner disk chemistry.
\subsection{Model set: External UV}\label{subsection: Ext. UV}
In the model set hereafter designated \texttt{Ext.\,UV}, we include an external FUV radiation field produced by OB stars in a rich stellar cluster, with an intensity of 10$^{4}$ $G_0$ in Habing units \citep{Habing_1968, Tielens_1985}. This value is representative of the average external UV irradiation measured for disks found in high-mass star-forming regions \citep{Winter_2022, Tannus_2025, Anania_2026}. The external FUV radiation field is implemented in ProDiMo as an additional isotropic background radiation field. Other than this, we keep the same star-disk structure of the \texttt{Baseline} set. 

In simple terms, as external FUV irradiation increases the gas temperature in the outer disk layers, the sound speed in the gas should also increase ($c_s\,\propto\,\sqrt{T}$), thus shrinking the gravitational radius deeper and deeper into the disk  ($R_{\rm g}$ = $GM_{\star}/c_s^2$), which defines where the gas becomes gravitationally unbound. The resulting flow is launched subsonically predominantly from the FUV-heated outer disk rim, with a contribution from the irradiated disk surface. This outflow will accelerate under its thermal pressure gradient, eventually becoming supersonic and expanding into a photoevaporative wind over an increasing solid angle \citep{Johnstone_1998, Winter_2022}. For the model set here considered, this implies that we are assuming that one or multiple OB stars have just switched on and have not removed significant material from the disk yet, or the disk is newly exposed to external irradiation after having left a region with high extinction.

As it can be seen in Fig. \ref{figure: ExtUV grid properties}, the major difference in disk structure compared to the baseline models consists in the formation of an irradiated wall at the outer disk radius ($\approx$100 au) that propagates inward in the upper disk layers, where a temperature inversion occurs and the disk remains warmer than usual (see $T_{\rm gas, dust}$ central panels in Fig. \ref{figure: ExtUV grid properties}). This is in agreement with previous modeling of externally irradiated T\,Tauri \citep{Portilla-Revelo_2025, Calahan_2025, Hernandez_Arboleda_2026}. Owning to the large absorption cross section of PAHs to UV photons, PAH heating takes over in a larger portion of disk, beyond 10 au, where the isotropic external FUV field well penetrates (see bottom left panel in \ref{figure: ExtUV grid properties}). PAHs will in fact absorb the OB star FUV flux, populating rovibrational states which then de-excite in collisions with with H$_2$ molecules that in turn warms up the gas.

In the outer disk, CO and H$_2$O roto-vibrational cooling also become less preponderant, as the increased intake of FUV photons pump molecules in higher excitation states, counteracting their line cooling. The opposite effect is observed for H$_2$ line cooling, which starts occurring also at the irradiated outer wall thanks to combination of increased gas temperature and efficient H$_2$ self-shielding.

\subsection{Model set: Truncated External UV}\label{subsection: Ext. UV with trunc.}
Finally, in the model set hereafter designated \texttt{Trunc.\,Ext.\,UV}, we account for the expected effects of external UV photoevaporation on the evolution of the gas component of planet-forming disks. To simulate the scenario of outer depletion and disk truncation in ProDiMo, we follow the approach introduced in \citet{Portilla-Revelo_2025} and \citet{Hernandez_Arboleda_2026} of imposing a smaller tapering-off radius, which here assumes the meaning of the radial distance at which the disk is truncated in response to the external FUV photoevaporation. We also impose a decrease of the gas mass beyond this tapering-off radius at fixed dust mass, in order to mimic the photoevaporative wind-like structure. Since disk truncation is a time-dependent effect, while our ProDiMo models are static, in this set of models we aim to simulate a snapshot in time of a planet-forming disk of increasing stellar host mass being truncated by external FUV photoevaporation to a certain size. 

More specifically, we built a two-zone ProDiMo setup where the innermost zone corresponds to the compact, truncated disk, and the outer zone to the disk region is photoevaporated. As for the \texttt{Ext.\,UV} model set, we simulate here the presence of an external FUV field equivalent to 10$^{4}$ $G_0$. The properties of the two zones, which are treated as physically and chemically different disk regions in ProDiMo, are set to be exactly the same except for the radii (inner and outer), the disk mass ($M_{\rm disk}$), and the dust-to-gas mass ratio ($d/g$). This modeling approach is equivalent to modifying the gas surface density profile in input in ProDiMo, as done in \citet{Portilla-Revelo_2025, Hernandez_Arboleda_2026}, but provides more control over the physical properties of the truncated disk structure. 
\begin{table*}
\centering
\caption{Mid-infrared gas-phase species and ProDiMo radiative transfer treatment included in the IRIS grid.}
\label{table: lines and rms}
\begin{tabular}{lcllc}
\hline
\hline
\noalign{\vskip 0.5mm}
Species & $N_{\rm lines}$ & ProDiMo treatment & Emission component &
Integrated flux region [$\mu$m]\\
\noalign{\vskip 0.5mm}
\hline
\noalign{\vskip 0.5mm}
CO & 939 & non-LTE \citep{Thi_2013} &
high-$J$ rovibrational band &
4.90--5.11\\
\hline
\noalign{\vskip 0.5mm}
CH$_3^+$ & 11659 & LTE \citep{Changala2023} &
$\nu_2$ rovibrational band &
7.10--7.20\\
\noalign{\vskip 0.5mm}
\hline
\noalign{\vskip 0.5mm}
C$_2$H$_2$ & 4313 & LTE \citep{gordon_2022} &
$\nu_5$ rovibrational band &
13.65--13.75\\
\noalign{\vskip 0.5mm}
\hline
\noalign{\vskip 0.5mm}
HCN & 4621 & non-LTE \citep{Bruderer_2015} &
$\nu_2$ rovibrational band &
13.95--14.06\\
\noalign{\vskip 0.5mm}
\hline
\noalign{\vskip 0.5mm}
CO$_2$ & 3692 & non-LTE \citep{Bosman_2017} &
$\nu_2$ rovibrational band &
14.90--14.99\\
\noalign{\vskip 0.5mm}
\hline
\noalign{\vskip 0.5mm}
H$_2$O & 14937 & non-LTE \citep{gordon_2022} &
[7 $\mu$m] line forest &
6.40--7.05\\
& &  &
[12 $\mu$m] line forest &
12.52--12.53\\
& & &
[17 $\mu$m] line forest &
17.08--17.39\\
& & &
[24 $\mu$m] line forest &
23.80--23.96\\
\noalign{\vskip 0.5mm}
\hline
\noalign{\vskip 0.5mm}
OH & 1023 & non-LTE \citep{Brooke_2016} &
$N=34$ line &
10.069--10.071\\
& & \citep{Tennyson_2024} &
$N=19$ line &
15.290--15.299\\
& & &
$N=18$ line &
16.015--16.025\\
\noalign{\vskip 0.5mm}
\hline
\noalign{\vskip 0.5mm}
H\,\textsc{i} & 189 & non-LTE \citep{Backs2023} &
H(10--6) line, $n=10\rightarrow6$ &
5.125--5.130\\
& & &
H(9--6) line, $n=9\rightarrow6$ &
5.901--5.910\\
& & &
H(6--5) line, $n=6\rightarrow5$ &
7.455--7.470\\
& & &
H(8--6) line, $n=8\rightarrow6$ &
7.495--7.510\\
& & &
H(7--6) line, $n=7\rightarrow6$ &
12.370--12.395\\
\noalign{\vskip 0.5mm}
\hline
\noalign{\vskip 0.5mm}
H$_2$ & 8215 & non-LTE \citep{Wrathmall_2007} &
H$_2$(S5) ortho-line, $J=7\rightarrow5$ &
6.909--6.911\\
& & \citep{woitke_2009} &
H$_2$(S4) para-line, $J=6\rightarrow4$ &
8.021--8.030\\
& & &
H$_2$(S3) ortho-line, $J=5\rightarrow3$ &
9.661--9.670\\
& & &
H$_2$(S2) para-line, $J=4\rightarrow2$ &
12.275--12.280\\
& & &
H$_2$(S1) ortho-line, $J=3\rightarrow1$ &
17.030--17.039\\

\hline
\end{tabular}
\begin{tablenotes}
    \footnotesize
    \item \textbf{Notes.} Column (1) indicates the atomic or molecular species. Column (2) and (3) respectively report number of included transitions and line treatment in ProDiMo. Column (4) states the nomenclature of the emitting component (line emission or rovibrational band) of the labeled species. Column (5) shows the spectral window used to estimate the integrated line flux of each emitting component in the synthetic spectra.
    \end{tablenotes}
\end{table*}

The inner zone in each model of this set is defined between the interpolated inner radius ($R_{\rm in}$ in Table~\ref{table: fixed parameters}) and a fixed tapering-off radius of 10 au. As there is no direct observational constraint mapping outer disk masses and radii to an external photoevaporation state, the choice of this radius is arbitrary. Nevertheless, this modeling approach, and this particular $R_{\rm tap}$ value, reproduces reasonably well the JWST observations of the XUE 1 irradiated disk discussed in \citet{Portilla-Revelo_2025}. Considering the customary $d/g$ = 10$^{-2}$, the disk mass in this zone ($M_{\rm disk, inner}$) is defined as the integral of the gas surface density profile $\Sigma_{\rm gas}$(r) of the correspondent reference \texttt{Baseline} model between $R_{\rm in}$ and 10 au, where generally:
\begin{equation}
 \Sigma_{\rm gas}(r) \propto r^{-\epsilon} \cdot \exp \left(-(r / R_{\rm tap})^{2 - \epsilon}\right)   
\end{equation}
and
\begin{equation}\label{equation: mass}
 M_{\rm disk} =  2 \cdot 2\pi \int \Sigma_{\rm gas}(r)~r~ dr 
\end{equation}
The surface density (in units of g cm$^{-2}$) is defined as a function of the radial coordinate $r$ with a power law shape of index $\epsilon$ \citep{Woitke_2016}, and is taken from the ProDiMo output of the \texttt{Baseline} set of models for each stellar mass. The factor of 2 in Eq. \ref{equation: mass} for the disk mass (in units of grams) takes into account for the symmetry of $\Sigma_{\rm gas}$ with respect to the disk midplane \citep{woitke_2009}. 

The outer zone is then defined between the truncated tapering-off radius of 10 au, and an outer radius $R_{\rm out}$ of 300 au. In order to simulate a gas-depleted outer zone, here we fix a dust-to-gas ratio of $d/g$ = 1, and we compute $M_{\rm disk,outer}$ as the leftover mass (total - inner), that is, the reservoir the outer zone could have had in absence of external FUV photoevaporation, diminished by a fixed depletion factor ($f_{\rm dep}$) of 1000. In this way, the outer disk mass still scales with each model's overall mass budget, but is uniformly suppressed relative to the inner disk. In other words:
\begin{equation}
    M_{\rm disk, outer} = \frac{M_{\rm disk, tot} - M_{\rm disk, inner}}{f_{\rm dep}}
\end{equation}
where $M_{\rm disk, tot}$ is the one defined in Table~\ref{table: structural parameters} for each stellar mass model, and $M_{\rm disk, inner}$ is the one integrated in the correspondent inner zone from Eq. \ref{equation: mass}. The single fixed value of $f_{\rm dep}$ provides a fairly uniform $\sim$100$\times$ outer disk depletion across the whole set of \texttt{Trunc.\,Ext.\,UV} models. This outer disk region is
therefore treated as a parametrized, gas-poor outer component rather than as a self-consistent external photoevaporative wind \citep{Keyte_2026}. Similarly to the choice of a truncation radius of 10 au, the value of disk mass in this zone is arbitrary due to the lack of robust observational constraints on different photoevaporative stages of externally irradiated disks.

The resulting disk structure from this model set is plotted in Fig. \ref{figure: Truncated grid properties} for two representative stellar mass models as shown for the rest of the IRIS grid. In presence of disk truncation, the thermochemical structure of the disk appears to be dominated by internal stellar irradiation (see e.g., $N_{\rm tot}$ panels in \ref{figure: Truncated grid properties}), thus resembling our baseline setup, although the dust stays warmer in the outer rarefied zone (see $T_{\rm dust}$ panels in Fig. \ref{figure: Truncated grid properties}). Moreover, with respect to the other model sets, the vertical column density of OH is predicted to remain nearly constant at $\sim\,10^{18}$ cm$^{-2}$ across the disk, extending to large radii, while the column densities of HCN and C$_2$H$_2$ decrease from $\sim$10$^{20}$ cm$^{-2}$ at 3 au to 5 au by respectively one order of magnitude and four orders of magnitudes, producing a radial inversion in the HCN-C$_2$H$_2$ ratio\footnote{As explained in \citet{Walsh_2015}, since the abundance of HCN is regulated by UV photochemistry via formation of H$_2$ and CN, and destruction via collisional dissociation, its emitting layer tends to be located deeper in the disk atmosphere at larger density, leading to a higher vertical column density compared to C$_2$H$_2$. This effect must be amplified by the fact that this model set assumes a luminosity in the UV range up to 40\% (for $M_{\star}$ = 4.0 $M_{\odot}$) of the total bolometric luminosity.}.

Heating and cooling mechanisms active in the disk (see bottom row of Fig. \ref{figure: Truncated grid properties}) remain relatively unvaried from the baseline models. The only exceptions are the viscous heating, which naturally spreads out following the extended density distribution, and the cooling channels via C\,II and O\,I atoms, which are favored at the high temperatures and low gas densities now present in the outer irradiated disk. The presence of FUV-induced truncation for disk radii $>$\,10\,au therefore generally counteracts the effects of external irradiation on the disk surface, resembling in the innermost disk regions the chemical gas-phase abundances of stellar-only irradiated models.

\section{Analysis methods}\label{section:Methodology}
 For the purposes of this work, we focus on a set of 9 chemical species: atomic hydrogen (H\,\textsc{i}), molecular hydrogen (H$_2$), carbon monoxide (CO), water (H$_2$O), hydroxide (OH), carbon dioxide (CO$_2$), acetylene (C$_2$H$_2$), hydrogen cyanide (HCN), and methyl cation (CH$_3^{+}$), which all emit in the gas-phase in the inner disk region at infrared wavelengths, traceable for example with JWST. In Sect. \ref{section: selection of disk observables} we justify the chosen sample of diagnostic species, and in Sect. \ref{section: line flux estimation} we describe how we derive line flux densities and line ratios taking into account spectral contamination from the abundant H$_2$O emission often present at mid-IR wavelengths.
\subsection{Selection of disk observables}\label{section: selection of disk observables}
Table \ref{table: lines and rms} contains the spectral ranges used to identify the main spectral feature(s) in emission of the chosen chemical species. These all represent crucial diagnostics for the structure and evolution of disks, as well as for the composition of forming planets. In particular, CO, H$_2$O, CO$_2$, C$_2$H$_2$, and HCN represent the fundamental carbon (C) and oxygen (O) molecular carriers in inner disks, therefore presumed to trace the C/O disk ratio. The infrared emission of CO, H$_2$O, and CO$_2$ is also connected to the
thermal and chemical disk structure around their respective snow lines, where
volatiles transition between the gas and ice phases and thereby redistribute
C- and O-bearing species between the gas and solid reservoirs
\citep[e.g.][]{Oberg_2011, Piso_2015}. Near snow lines, these gas-phase molecules can also alter dust evolution by affecting grain sticking and fragmentation and enhancing the local density of icy solids through vapor diffusion and recondensation, hence promoting planetesimal formation.
\citep[e.g.][]{Cieza_2016,Schoonenberg_2017,Drazkowska_2017}. Hydrogen and OH emission lines are instead proxies of the ionization state of the disk, while C$_2$H$_2$ and HCN have shown to be sensitive to the input of X-ray radiation impinging on the disk \citep{Woitke_2024_EXLupi}. CH$_3^{+}$, specifically, has been connected to the presence of a UV irradiated disk cavity, but also to external UVs impinging on the disk surface \citep{Henning_2024, Berne_2023}.
 
 We exclude from our investigation the emission lines coming from ionized species like neon (Ne\,II, Ne\,III) and argon (Ar\,II, Ar\,III). These are well-known diagnostics for the presence of UV and X-ray photoevaporative winds at the disk surface, while the thermal disk component contributing to their emission lines is relatively low \citep{Glassgold_2007, Meijerink_2008, Ercolano_2008, Ercolano_2010, Schisano_2010}. Although the analysis of these species would be compelling for the purposes of this work, our ProDiMo simulations do not explicitly model a physically distinct wind region (see also Sect. \ref{discussion: model limitations}). Hence, we can only take into account the thermal contribution to these lines, which would strongly underestimate the real emission of these species in the disk. Even though H$_2$ itself can be associated with outflow emission \citep[e.g.,][]{Nakatani_2026}, we include in this work the analysis of the most commonly observed and brightest H$_2$ mid-infrared ortho-para S(1)--S(5) lines (see Table~\ref{table: lines and rms}), which are good disk diagnostics as a function of UV irradiation, luminosity, and accretion of the host star \citep[e.g.,][]{Thi_2001, Carmona_2008, woitke_2009}.
 \begin{figure*}[!htbp]
\centering
    \includegraphics[width=\textwidth]{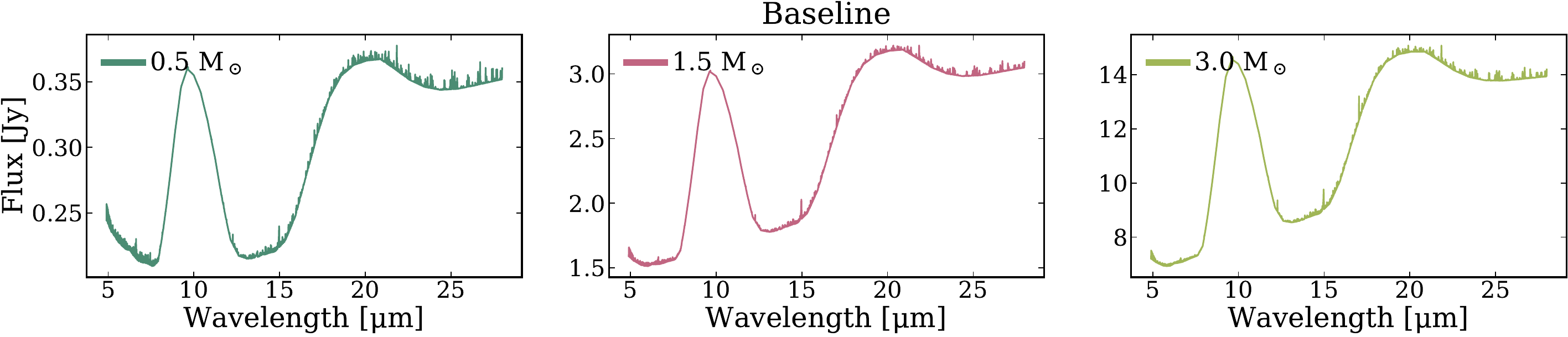}
    \caption{Synthetic ProDiMo mid-IR spectra from the \texttt{Baseline} set of models for increasing stellar host mass. The spectra are generated with FLiTs in LTE and non-LTE (see Table~\ref{table: lines and rms}), convolved to a JWST/MIRI resolution of 2680 (MRS at $\sim$\,15 $\mu$m), and scaled for convenience to a distance of 140 pc, typical of nearby isolated star-forming regions.}
    \label{figure: prodimo spectra baseline}
\end{figure*}
\begin{figure*}[!htbp]
\centering
\includegraphics[width=\textwidth]{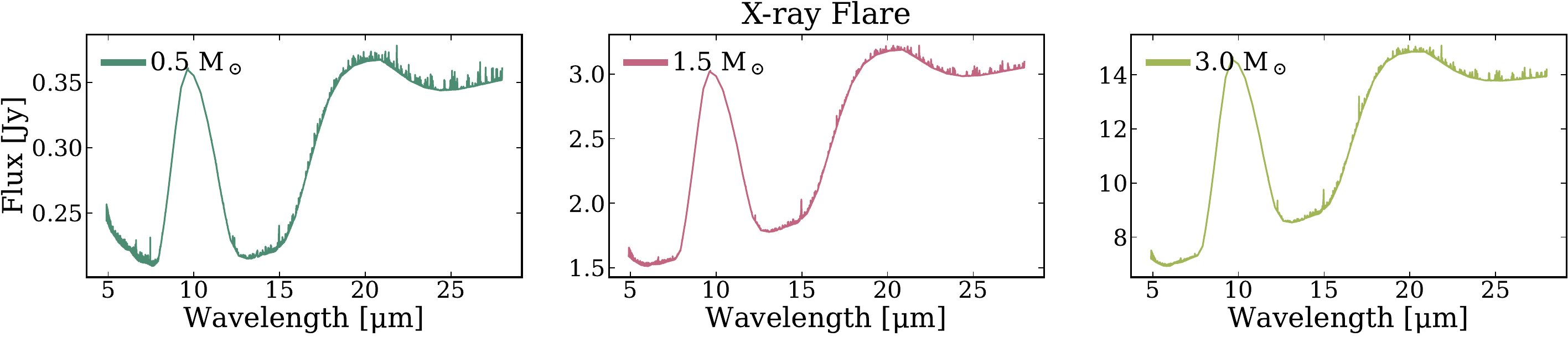}
    \caption{Same as Fig. \ref{figure: prodimo spectra baseline} but for the \texttt{X-ray\,Flare} set of models.}
    \label{figure: prodimo spectra xray flare}    
\end{figure*}
\begin{figure*}[!htbp]
\centering
    \includegraphics[width=\textwidth]{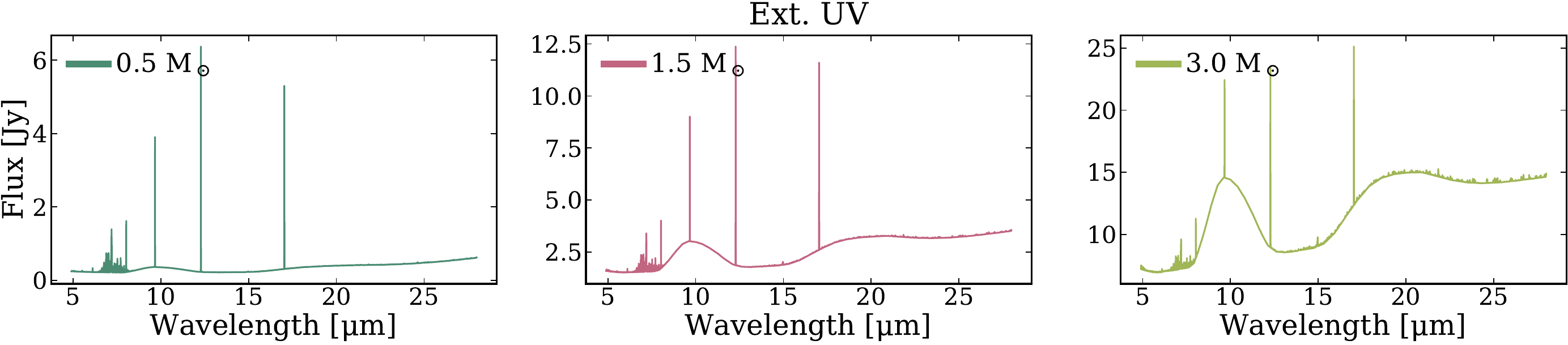}
    \caption{Same as Fig. \ref{figure: prodimo spectra baseline} but for the \texttt{Ext.\,UV} set of models.}
    \label{figure: prodimo spectra extuv}    
\end{figure*}
\begin{figure*}[!htbp]
\centering
    \includegraphics[width=\textwidth]{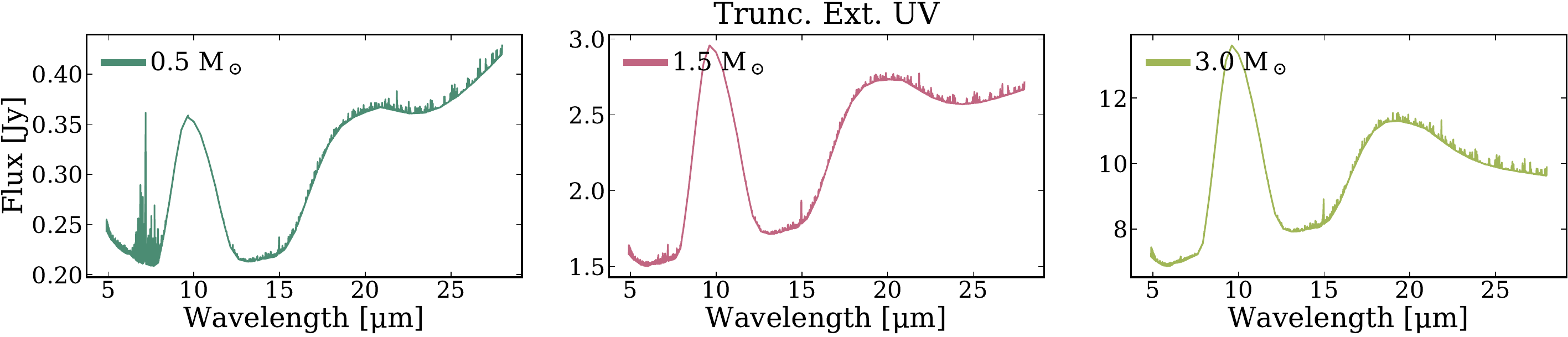}
    \caption{Same as Fig. \ref{figure: prodimo spectra baseline} but for the \texttt{Trunc.\,Ext.\,UV} set of models.}
    \label{figure: prodimo spectra truncated}    
\end{figure*}
 
\subsection{Line flux estimation}\label{section: line flux estimation}
We proceed by computing the integrated flux ($F_{\rm int}$) of our chosen set of species diagnostics in the spectral windows where their emission is mainly expected to arise (see Table~\ref{table: lines and rms}). This consists of either single, unblended emission lines, e.g., the OH lines identified by the rotational quantum number $N$ = 34, $N$ = 19, $N$ = 18 (hence \enquote{OH(N34)}, \enquote{OH(N19)}, \enquote{OH(N18)}), or roto-vibrational bands, e.g., the H$_2$O forest of rovibrational lines expected around 6--7 $\mu$m (hence \enquote{H$_2$O [7 $\mu$m]}). 

We numerically sum the positive contributions to the flux density within the wavelength range of the features of interest. In order to obtain reliable flux measurements of individual species, we assess the incidence of water lines overlapping with the emission coming from other molecular species of our interest in this work, i.e., potentially contaminating our flux estimates for those species. In our set of selected emission lines (see Table~\ref{table: lines and rms}), we identify the rovibrational bands of CO, CH$_3^+$, C$_2$H$_2$, HCN, CO$_2$, and some lines of atomic hydrogen (identified by a change in principal quantum number $\Delta n$ = 6--5, 8--6, 7--6) and of molecular hydrogen (S(2), or S2, identified by a change in the rotational quantum number by +2, ($\Delta J$ = 4--2) as potentially contaminated by water emission. We verified this by looking at the radiative transfer slab models \citep[in 0D LTE;][]{Tabone_2023} for the expected emission of each species in the MIRI range. 

The mid-IR spectra generated with FLiTs in ProDiMo consist of blended lines of the atomic and molecular species emitting in the wavelength range given in input by the user. In order to separate and remove the contributing flux of water from the emission coming from the rest of the species, we created and subtracted a FLiTs spectrum of water from the FLiTs spectrum including all species blended with water. We applied this approach to all stellar mass models across the IRIS grid. We emphasize that this method for the assessment of spectral contamination is only approximate. The full radiative transfer solution cannot be recovered by simply adding the water spectrum to the spectra of the remaining species at common wavelengths, as line emission couples to both the dust continuum and overlapping molecular opacities, altering the emergent spectrum.

The results of our assessment are summarized in Fig. \ref{figure: H2O contamination}. Water contamination is estimated as the fraction of the total integrated flux that is due to water emission. Defined $F_{\rm int,\,tot}$ as the integrated flux density with water included, and $F_{\rm int,\,sub}$ as the integrated flux where the contribution from water is subtracted, we estimate water contamination as [($F_{\rm int,\,tot}$--$F_{\rm int,\,sub}$)/$F_{\rm int,\,tot}$] $\times$\,100. A percentage of 100\% would therefore indicate a wavelength region completely dominated by water emission.
At fixed species, the different sets of models predict similar ranges of water contamination. CO, HCN, CO$_2$, and H$_2$(S2) are the least contaminated emitting features, with a flux contamination spread around or below 60\%. As expected, instead, CH$_3^+$ is highly affected, due to the presence of relatively bright water rovibrational lines in the 7.1--7.3 $\mu$m range, overlapping with the main rovibrational $\nu_2$ band of CH$_3^+$. The low contamination predicted by the \texttt{Ext.\,UV.} set is related to the fact that the contrast between CH$_3^+$ and H$_2$O is expected to be the largest among the IRIS grid; see also Sect. \ref{section:Results}. The same argument applies to the atomic hydrogen recombination lines, which are widely used as a diagnostic of stellar accretion and are known to be frequently contaminated by water emission \citep[e.g.,][]{Tofflemire_2025, Shridharan_2026}. 

The line fluxes of the contaminated species are therefore computed in the synthetic spectra minus the flux from the H$_2$O only FLiTs spectra, while for the other species they are measured in the full FLiTs output spectra. The only exception is the integrated flux of C$_2$H$_2$ in the $M_{\star}$ = 4.0 $M_{\odot}$ model across all sets, which was measured in the original spectra due to the production of artifacts from water subtraction against the very bright dust continuum. 

The final integrated line fluxes are visualized altogether for all sets of models in Fig. \ref{figure: line luminosities versus mass}, and are discussed in detail in the following Section.

\section{Results}\label{section:Results}
The new set of ProDiMo models built with our grid allows to analyze the effects of internal and external disk irradiation on mid-IR (4.9--28 $\mu$m) spectra and gas-phase emission of inner disks surrounding typical \textit{FGKM}-type 0.3--4.0 M$_\odot$ pre-main sequence stars, particularly at the sensitivity and resolution of JWST.

\begin{figure*}[htp!]
    \centering
    \includegraphics[width=0.88\linewidth]{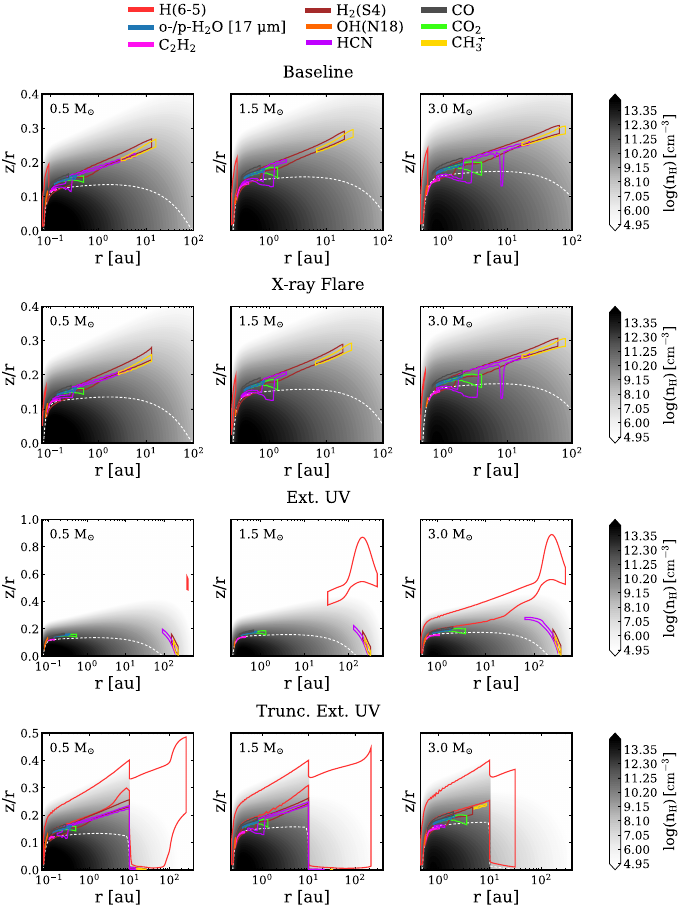}
    \caption{Emitting areas (colored boxes) tracing the origin in the disk along the radial (\textit{x}-axis in log$_{\rm 10}$ scale) and vertical direction (\textit{y}-axis, dimensionless) of the rovibrational mid-IR emission of selected volatile gas-phase species (see legend) included in the IRIS grid. Models are plotted with increasing stellar host mass from left to right. The mid-IR transitions shown are the brightest emitting lines identified in ProDiMo within each species flux integration region listed in Table \ref{table: lines and rms}. The plotted 17 $\mu$m water component refers to ortho (o-) and para (p-) molecules. The grayscale background shows the gas density structure of each model in units of hydrogen nuclei number density. The dashed white curve indicates a visual extinction of unity ($A_{\rm V}$ = 1), where the dust becomes optically thick to the emerging radiation.}
    \label{figure:emitting areas}
\end{figure*}

Figures \ref{figure: prodimo spectra baseline} to \ref{figure: prodimo spectra truncated} show the FLiTs mid-IR full spectra simulated at JWST/MIRI MRS-like resolution according to the different star-disk irradiation setups explained in Sect. \ref{section:Model setup} for a pre-main sequence star of mass M$_\star$ = 0.5 M$_\odot$, 1.5 M$_\odot$, and 3.0 M$_\odot$ (from left to right in the figures), scaled to 140 pc. The adopted distance scaling is a normalization choice, but facilitates direct comparison with fluxes typically observed for disks in nearby isolated star-forming regions such as Taurus. Figures \ref{figure: baseline spectra zoom}--\ref{figure: truncated spectra zoom} zoom on the continuum-subtracted spectra in windows defined between 4.9--5.2 $\mu$m, 13.4--15.8 $\mu$m, and 20--25 $\mu$m, which cover relevant spectral features of some of the species analyzed in this work. Figure \ref{figure:emitting areas} instead shows the 2D disk regions from which the mid-IR emission of the analyzed species in the IRIS grid spectra originates. In the case of  H\,\textsc{i}, H$_2$ and H$_2$O, we only plot their brightest emitting components, or the ones which are the most commonly used disk diagnostics (see legend in figure).

The integrated fluxes computed with the IRIS grid are plotted as a function of stellar mass in Fig. \ref{figure: line luminosities versus mass}. The general predicted trend is that of increasing line flux density with increasing stellar host mass \citep{Walsh_2015, Antonellini_2015, Antonellini_2016}. This is also reflected by the emitting areas of the different species in the disk, which tend to be more extended in radius or in height in the higher stellar mass models across the grid (3.0 $M_{\odot}$ in Fig. \ref{figure:emitting areas}). The rovibrational emission visible in the mid-infrared windows considered in this work is mostly confined within 10 au (for CO, H$_2$O, OH, CO$_2$, C$_2$H$_2$), while in the outer disk in the FUV externally irradiated models H\,\textsc{i}, H$_2$, HCN and CH$_3^+$ mid-IR lines have contributions or are only emitted from low-density layers of the outer disk (third and bottom rows in Fig. \ref{figure:emitting areas}). 

The IRIS grid further predicts that the brightest disks in dust continuum and gas lines are those which are extended and externally FUV illuminated (Fig. \ref{figure: prodimo spectra extuv} and \ref{figure: extUV spectra zoom}), in agreement with previous modeling of external FUV irradiation of disks \citep{Walsh_2013, Calahan_2025, Portilla-Revelo_2025, Keyte_2026, Hernandez_Arboleda_2026}. This occurs because the external FUV field contributes to illuminating the whole disk surface, increasing the dust and the gas temperatures in the upper ($z/r$\,$>$\,0.2) and outer ($r$\,$>$\,10 au) layers, and, in particular, triggering more molecules in the gas towards excited states (see also Fig. \ref{figure: ExtUV grid properties}). External FUV irradiated disk models also display distinctive strong molecular H$_2$ and CH$_3^+$ line emission, except when the disk is truncated (Fig. \ref{figure: prodimo spectra truncated}). The results concerning each individual model set are presented more in detail in the paragraphs below, in order, for the reference baseline models (Sect. \ref{results: baseline}), X-ray flared (Sect. \ref{results: x-ray flaring}), and the externally FUV irradiated models (Sect. \ref{results: fuv irradiation}).

\subsection{Baseline models}\label{results: baseline}
Our starting point, represented by the \texttt{Baseline} model set (Sect. \ref{subsection: Baseline} and Fig. \ref{figure: prodimo spectra baseline}), shows that we can qualitatively reproduce the line spectral richness observed in the mid-infrared region towards the majority of low-mass disk sources \citep[e.g.,][]{Henning_2024, Arulanantham_2025}. For higher stellar masses ($M_\star$\,$>$\,1.5\,$M_\odot$) disks are expected to be richer in molecular lines than the lower-mass counterparts, and progressively brighter (Fig. \ref{figure: baseline spectra zoom}), with most of the species (H\,\textsc{i}, H$_2$, CO, H$_2$O) always emitting from optically thin dust layers in the disk (first row in Fig. \ref{figure:emitting areas}). This is somewhat in contrast to the evidence from the infrared surveys of isolated disk-bearing Herbig Ae/Be stars done with various space- and ground-based instruments \citep[][]{Brittain_2007, Pontoppidan_2010, Salyk_2011, Fedele_2011, Meeus_2012, Ilee_2014, Brittain_2016, Banzatti_2018, Banzatti_2022, Banzatti_2023}, where most of them are reported with non-detections or faint molecular gas-phase emission. A variety in the volatile inventory is seen instead among the sample of externally irradiated XUE Herbig Ae/Be and IMTT disks observed toward NGC 6357 with JWST/MIRI \citep{Tannus_2025}, and, always with the same instrument, in the nearby isolated HD 35929 Herbig Ae/Be disk \citep{Kaufer_2026}. A detailed comparison of the IRIS grid with high sensitivity JWST infrared spectra of both T\,Tauri and Herbig Ae/Be disks, and a discussion on modeling aspects that could explain the appearance of line-poor Herbig Ae/Be spectra, is presented in a separate paper (Frediani et al., in prep).

Baseline models and those including an external FUV field (with or without truncation) or a X-ray stellar flare yield a similar range of line fluxes over the entire stellar mass range for species like CO, C$_2$H$_2$, HCN, CO$_2$, H$_2$O and OH (Fig. \ref{figure: line luminosities versus mass}). The exception is represented by the rovibrational emission component of water arising in the 6--7 $\mu$m range (\enquote{H$_2$O [7 $\mu$m]} in Fig. \ref{figure: line luminosities versus mass}), whose flux density is predicted to be higher in truncated and externally irradiated T\,Tauri disks. The apparent enhancement of this H$_2$O band in our low-mass models is due to contamination from the hot-band of CH$_3^+$, peaking at 7.1 $\mu$m outside our H$_2$O band, rather than real increased H$_2$O emission. In fact, CH$_3^+$ emission is strongly enhanced with respect to water in presence of external FUV irradiation, respectively by a factor$\sim$50 in irradiated models with full disks, and $\sim$8 in truncated models (Sect. \ref{results: fuv irradiation}). In the Herbig Ae/Be regime, \texttt{Ext.\,UV} models remain brighter in CH$_3^+$ by a factor$\sim$4 than H$_2$O at these wavelengths. In the corresponding truncated models, as truncation removes a substantial part of the warm, FUV-illuminated emitting disk surface, CH$_3^+$ gets fainter and its contrast with H$_2$O negligible. In fact, the truncated Herbig Ae/Be models retain only about 1\% of the original projected emitting area of CH$_3^+$ emission, whereas the truncated T-Tauri models retains roughly 9\% (Fig. \ref{figure:emitting areas}).

\subsection{Stellar X-ray flaring}\label{results: x-ray flaring}
The inclusion of a stellar X-ray flare has negligible effects on the line-to-continuum ratio and on the line brightness of the gas species analyzed in this work with respect to the baseline models, as also evident from their emitting areas (second row in Fig. \ref{figure:emitting areas}). Only atomic hydrogen lines seem to be slightly enhanced (by a factor $< 2$) in response to a higher X-ray luminosity, especially the H(6--5), H(8--6), and H(7--6) lines arising between 7 and 12 $\mu$m (Fig. \ref{figure: prodimo spectra xray flare} and corresponding panels in Fig. \ref{figure: line luminosities versus mass}).

The absence of strong X-ray flare-driven signatures could be due to several factors. First, the super-flare-like X-ray luminosity that we implement in the IRIS grid ($\log(E_{\rm X}) \sim 34$--35 erg) may simply not be energetic enough to produce a detectable effect. Our models do not address the much rarer but substantially more energetic X-ray mega-flares ($\log(E_{\rm X}) \sim 36$--38 erg), which could plausibly leave a stronger imprint (Getman et al., in prep.). Second, we may not have chosen the proper mid-infrared diagnostic species or wavelength window to expect a pronounced difference. As mentioned in Sect. \ref{section: selection of disk observables}, we exclude from our analysis diagnostics such as [Ne\,II] and [Ne\,III], for which X-ray flares could make a more noticeable difference (Portilla-Revelo et al., in prep.). Third, our treatment of X-ray flares is admittedly simplified: X-ray photon reprocessing through scattering and absorption could reveal flare-induced chemical and spectroscopic effects that our current treatment misses, since in our models X-rays do not efficiently penetrate the disk. 

\subsection{External FUV irradiation}\label{results: fuv irradiation}
\begin{figure*}[!htbp]
\centering  \includegraphics[width=\textwidth]{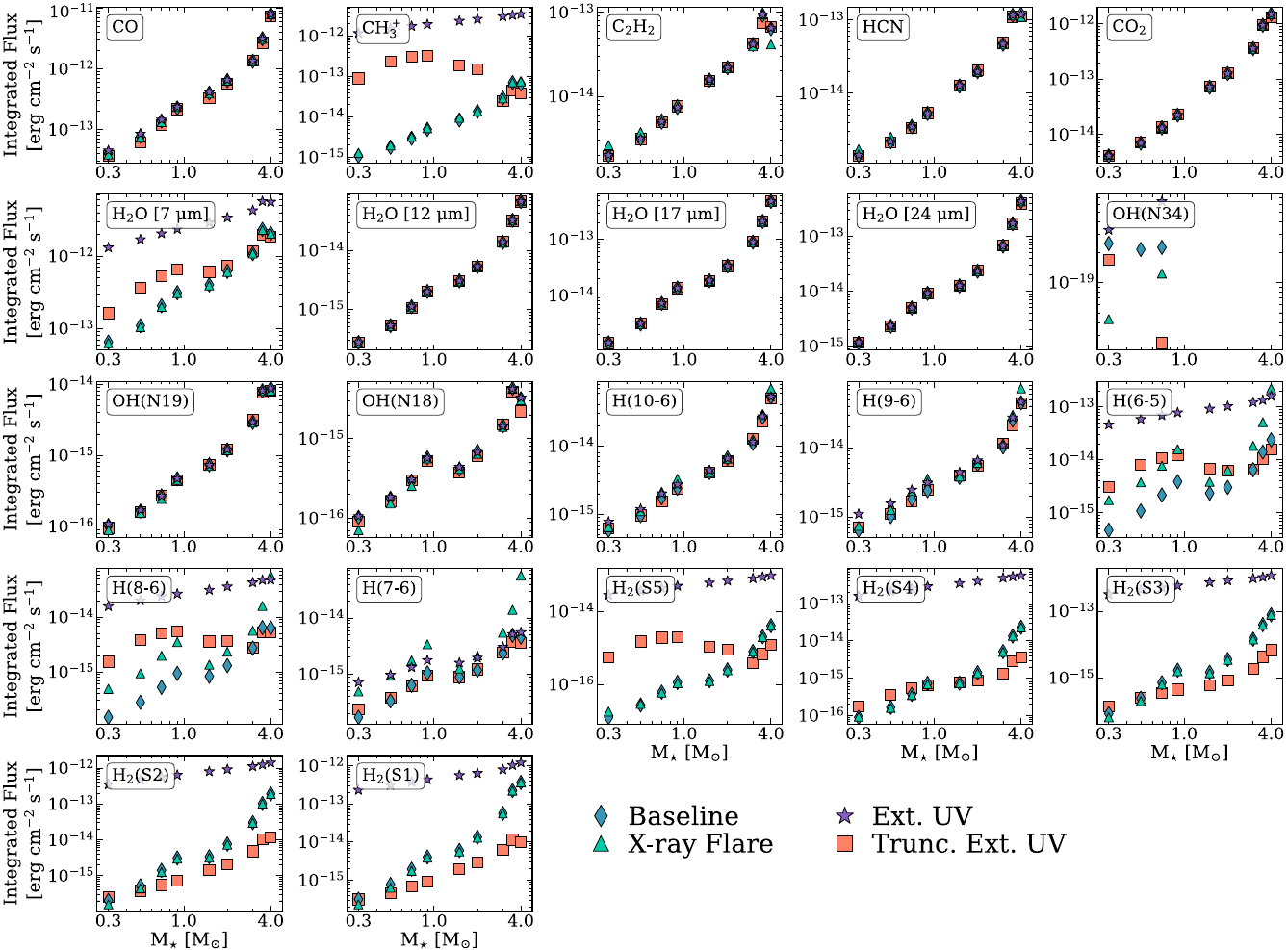}
    \caption{Predicted mid-infrared fluxes (scaled to 140 pc) from the IRIS grid as a function of stellar mass in log-log scale for the emission components of the species analyzed in this work, integrated over the wavelength ranges indicated in Table~\ref{table: lines and rms}.}
    \label{figure: line luminosities versus mass}   
\end{figure*}

By adding an external FUV Habing field of 10$^4$ $G_0$ in the star-disk setup (with no truncation), the expected outcome in the mid-IR spectral range is the appearance of the already mentioned bright H$_2$ line emission, and of a strong signal from the CH$_3^+$ rovibrational band centered around 7 $\mu$m (Fig. \ref{figure: prodimo spectra extuv}). While the dust continuum becomes brighter, the line emission from other species (e.g., CO, H$_2$O, CO$_2$, C$_2$H$_2$, HCN) remains comparable to that of our \texttt{Baseline} (or \texttt{X-ray\,Flare}) models, independently of stellar mass. Overall, externally illuminated disk models without truncation predict systematically higher line fluxes by $\sim$one to two orders of magnitude in the case of CH$_3^+$, H\,\textsc{i} (with $\Delta n = 8$--6 and $\Delta n = 6$--5), and of H$_2$ emission lines than any other set of models (Fig. \ref{figure: line luminosities versus mass}).

A careful inspection of the vertical integrated emission (weighted by dust optical depth) computed with ProDiMo reveals that the enhanced H$_2$ and CH$_3^+$ rovibrational emission arise from relatively cold outer disk molecular layers (third row in Fig. \ref{figure:emitting areas}). In particular, the H$_2$ emission is associated with a gas temperature of $\sim$415 K and a column density $> 10^{20}$ cm$^{-2}$ (both for molecules in ortho and para states), emitting at a disk radius ranging between 155 au (M$_\star = 0.3$ M$_\odot$) and 304 au (M$_\star = 4.0$ M$_\odot$). This behavior is consistent with the efficient photoelectric heating on dust grains (or PAHs) by the additional external UV field, which maintains an extended layer of warm, self-shielded molecular gas in the disk surface and substantially enhances the H$_2$ rotational emission (Fig. \ref{figure: ExtUV grid properties}).

Meanwhile, CH$_3^+$ seems to emit at a slightly higher temperature of $\sim$440 K in a disk emitting radius of $\sim$155--260 au at $< 10^{13}$ cm$^{-2}$. Unlike other molecular species that are regulated by thermal desorption from dust grains, the abundance of CH$_3^+$ is controlled by ion--molecule chemistry in the gas. In particular, its main formation channel proceeds through the hydrogenation of the CH$_3^+$ ion, whose production can be enhanced in UV-irradiated environments by reactions involving vibrationally excited H$_2$ and C$^+$ ions \citep[e.g.,][]{Agundez_2010, Zannese_ions}. Consequently, CH$_3^+$ is expected to trace warm, photodissociation region (PDR)-like surface layers, making our inferred excitation temperature, emitting radii and column densities physically plausible. However, since CH$_3^+$ excitation in such low-density UV-irradiated layers is unlikely to be fully thermalized, LTE may not strictly apply. Hence, our derived temperature should be regarded only as an excitation temperature rather than a measurement of the gas kinetic temperature.

By cutting and depleting the outer gas disk in the \texttt{Trunc.\,Ext.\,UV} models, both the CH$_3^+$ and the H$_2$ line emission get suppressed by one to two orders of magnitude, with stronger suppression occurring in higher stellar mass models, for $M_{\star}$\,$>$\, 0.9 M$_{\odot}$. A similar phenomenon with CH$_3^+$ was found by \citet{Portilla-Revelo_2025} for the irradiated XUE 1 disk, where a truncated gas-depleted ProDiMo disk model better agrees with the absence of CH$_3^+$ in XUE 1's JWST spectrum. Nevertheless, truncated models retain brighter molecular line emission than non-irradiated models in CH$_3^+$, and H$_2$ with $\Delta J = 7$--5 (S5), but also H emission with $\Delta n$ = 6--5, 8--6, although the flux difference gets diluted with increasing stellar mass. In the case of H$_2$ lines characterized by $\Delta J = 3$--1 (S1), $\Delta J = 4$--2 (S2), $\Delta J = 5$--3 (S3), and $\Delta J = 6$--4 (S4), the predicted fluxes for truncated models tend to be lower than any other set of models, as it can also be noted comparing emitting areas (bottom row in Fig. \ref{figure:emitting areas}).

For the OH line with upper-state rotational quantum number $N$ = 34 (second row in Fig. \ref{figure: line luminosities versus mass}), our models behave unexpectedly with respect to the increasing line trends predicted for the other species, including the other OH lines included in this work, characterized by $N$ = 19 and $N$ = 18. In particular, our models systematically predict a very low OH(N34) flux density, or the line is seen in absorption instead of emission. This result can be largely explained by optical depth effects in our models. With increasing stellar mass, we observe in the IRIS grid that the 2D abundance of OH peaks in a warm molecular layer below the 10 $\mu$m dust photosphere, but the observed continuum originates from hotter silicate-emitting dust above it. This OH line therefore removes photons from the continuum emission, appearing in absorption in the spectra.

Secondly, high upper-level energy OH lines, e.g., $N$ = 34, require hotter and denser gas to be excited. These lines can therefore appear only when collisions can populate those upper levels efficiently, unless a pumping mechanism, such as radiative pumping, is present \citep[e.g.,][]{Woitke_2011, Tabone_2021, Zannese_2024}. Lower upper-level N lines, on the other hand, require less energy to be excited, and can therefore be easily populated over a much broader range of disk radii and heights, even without pumping mechanisms. In Fig. \ref{figure:emitting areas}, the low-energy OH emission line with $N$ = 18 is excited in a relatively vertically extended layer close to the inner disk rim. In our grid, since our non-LTE OH model does not include pumping (see Table \ref{table: lines and rms}), the $N$ = 34 population of OH molecules must be maintained purely collisionally. This explains, combined with the previous argument on the dust shielding, why the measurable OH mid-infrared emission at 10 $\mu$m remains very faint or is seen in absorption in the IRIS grid.

\section{Discussion}\label{section: Discussion}
The mid-infrared spectra and line fluxes emerging from the IRIS grid addresses the possibility of disentangling effects of the external environment from internal stellar radiation using observable disk properties (Sect. \ref{subsection: Disentangling external photoevaporation from infrared disk spectra}). Elemental ratios (C/O, O/H, and C/H) traced by molecular carriers in the inner disk and their dependency on stellar mass and disk irradiation can also be examined (Sect. \ref{section: C/O ratio}).

\subsection{Observational signatures of external photoevaporation in infrared disk spectra}\label{subsection: Disentangling external photoevaporation from infrared disk spectra}
Given the appearance of the different modeled grid spectra (Fig. \ref{figure: prodimo spectra baseline}--\ref{figure: prodimo spectra truncated}), it is possible to identify discriminants of externally irradiated disks from non-irradiated ones dominated by stellar irradiation, to be tested by current and future observations. The model spectra of externally FUV illuminated disks, with their distinctive bright CH$_3^+$ and H$_2$ emission features, resemble the photochemistry-dominated spectra of low-mass young proplyd objects observed in the high-mass Orion Nebula Cluster (ONC) with JWST \citep[e.g.,][]{Berne_2023, Schroetter_2025}. This suggests that at early times, with no ongoing severe gas or dust depletion, as currently suggested to be occurring in the ONC region \citep{Boyden_2023}, external FUV photoevaporation does leave an imprint on the disk infrared emission. We also show that this should be visible both in the T\,Tauri and Herbig Ae/Be domain, independently of the internal stellar UV field. 

In contrast, the IRIS simulated spectra do not provide clear distinctions for externally irradiated disks when they have already been significantly truncated. The absence of CH$_3^+$ booming emission, and the under-predictions of H$_2$ line fluxes at 8--17 $\mu$m from our truncated model set could be used as new age estimates for externally photoevaporating systems. The JWST/MIRI sample of XUE disks in NGC 6357 \citep{Tannus_2025}, on average a harsher irradiation environment than the ONC ($\gtrsim$\,10$^{5}$ $G_0$, priv. comm.), provide evidence that external FUV photoevaporation can truncate the disk and make its inner gas-phase features resemble non-irradiated sources. In fact, among all XUE disks there is only one tentative detection of CH$_3^+$ (Lemus et al., in prep.), whereas H$_2$ emission is fainter in comparison with lower-mass counterparts \citep{Tannus_2025}. 

As pointed out by \citet{Calahan_2025} and \citet{Keyte_2026} for a T\,Tauri disk model, only by considering very high external FUV fields ($\gtrsim$10$^{5}$ $G_0$) the increase in dust and gas temperatures is sufficient to significantly enhance gas-phase molecules like CO, H$_2$O and OH in the inner disk regions. Although the order of magnitude of the external FUV field set in the IRIS grid is more representative of the average value found in the known population of irradiated protoplanetary disks \citep{Winter_2022, Tannus_2025, Anania_2026}, our models may therefore sit in a regime where the external FUV field is not yet strong enough to drive dramatic changes in the inner disk chemistry of disks surrounding either T\,Tauri and Herbig Ae/Be stars. Adopting a 10$^5$ $G_0$ external FUV field in their ProDiMo modeling, \citet{Hernandez_Arboleda_2026} show that also C$_2$H$_2$ and HCN should be brighter in the JWST/MIRI range than in a non-irradiated T\,Tauri disk, assuming a tapering radius of 70 au, and the same PAH fraction adopted in this work ($f_{\rm PAH}$ = 10$^{-2}$). 

By truncating their disk model from 70 au to 15 au and 5 au, they also find that the line-to-continuum ratio of C$_2$H$_2$ and HCN should be strongly damped, up to 90\% for C$_2$H$_2$. This is not reflected in our 10 au truncated models with stellar host mass $M_{\star}$\,$<$\,1.5 $M_{\odot}$, where we find fairly constant integrated fluxes of HCN and C$_2$H$_2$ with respect to the stellar-only irradiated models, respectively of $\sim$5 $\times$ 10$^{-{15}}$ erg cm$^{-2}$ s$^{-1}$ (HCN), at least one order of magnitude fainter than \citet{Hernandez_Arboleda_2026}, and of $\sim$7 $\times$ 10$^{-{15}}$ erg cm$^{-2}$ s$^{-1}$ (C$_2$H$_2$), which is one to three orders of magnitude fainter compared to their tapered disk models. 

The lower mid-IR line fluxes of C$_2$H$_2$ and HCN observed in our models may arise from two different factors. First, by explicitly accounting for the line and the dust continuum opacity along a line of sight in the disk, the FLiTs post-processing included in the IRIS grid naturally tends to predict lower line fluxes than the escape probability formalism \citep[see also][]{Kanwar_2024}. In comparison, \citet{Hernandez_Arboleda_2026} ProDiMo models were computed without FLiTs and before the revision of the escape-probability treatment described in \citet{Woitke_2024_EXLupi}, where it is found that the old escape probability formalism can overestimate mid-IR line fluxes by
$\sim$\,10--30\% relative to the revised treatment. This implies that our models are likely not underestimating C$_2$H$_2$ and HCN line fluxes, but they are rather computed with a more physically complete treatment in FLiTs. Second, while we cannot evaluate excitation effects of C$_2$H$_2$ transitions, whose level populations are still treated solely in LTE in ProDiMo so far (Table~\ref{table: lines and rms}), differently than \citet{Hernandez_Arboleda_2026} we implement the new non-LTE line list of HCN (an order of magnitude larger than its LTE counterpart), and for which non-LTE excitation effects may be important \citep{Bruderer_2015}. In particular, the gas densities in the outer disk (a few tens of au) can fall below the critical density for many of HCN transitions. Indeed, in our grid only $\sim$\,5\% of them reach densities exceeding their critical densities, meaning that most of HCN transitions in our models emit in non-LTE conditions. Therefore, they are naturally predicted to have fainter fluxes relative to the LTE computations presented in \citet{Hernandez_Arboleda_2026}.

\subsection{Predicted abundance ratios in disk atmospheres}\label{section: C/O ratio}
\begin{figure}[!htbp]
    \centering
    \includegraphics[width=\linewidth]{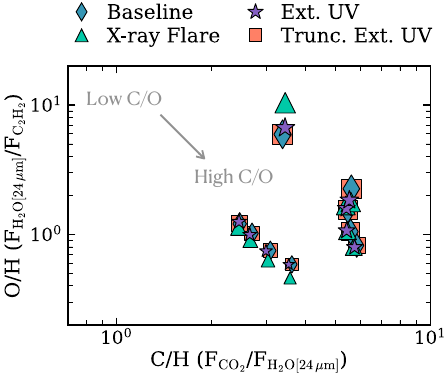}
    \caption{C/O ratio diagnostics from synthetic spectra of the IRIS grid. The \textit{x}- and \textit{y}-axis respectively show the gas-phase C/H and O/H ratios from line integrated fluxes of CO$_2$ (14.9 $\mu$m), H$_2$O (24 $\mu$m), and C$_2$H$_2$ (13.7 $\mu$m), according to Table~\ref{table: lines and rms}. Each set of models is coded by increasing marker size according to the assumed stellar mass. The grey arrow shows the direction of variation of the bulk C/O ratio modelled by \citet{Arabhavi_2026}.}
    \label{figure: C/O ratio}
\end{figure}

Having established the diagnostic potential of mid-infrared molecular emission across the IRIS grid, we now consider the C/O ratio and how it can be probed through disk irradiation.

We must recall that the bulk elemental C/O ratio describes the total carbon and oxygen content of the disk. In thermochemical models, this depends on the carbon and oxygen inventory available to the chemical network, and is fixed in our models at the adopted solar elemental value C/O = 0.457 (in Table \ref{table: fixed parameters}). By contrast, the local gas-phase C/O ratio can vary throughout the disk as carbon and oxygen partition differently between the gas and solids, for example through freeze-out and desorption on the surface of dust grains. Therefore, the C/O ratio inferred with molecular line ratios from infrared disk spectra is first of all an observational proxy for the gas-phase C/O ratio in the emitting layers of the molecules used to compute line ratios, which coincide with the upper warm layers of the disk atmosphere (Fig. \ref{figure:emitting areas}). Assuming that gas and dust are well mixed in the inner disk by turbulence, molecular line ratios are diagnostic of the fraction of bulk C/O ratio in disk atmospheres available for accretion onto protoplanets. 

This has been recently validated using H$_2$O, C$_2$H$_2$, and CO$_2$ as molecular tracers in the ProDiMo modelling work of \citet{Arabhavi_2026}, who show that combinations of their mid-IR JWST/MIRI line flux ratios can trace the elemental C/H and O/H model abundances adopted in their models, and therefore constrain the underlying elemental bulk C/O ratio. Figure~\ref{figure: C/O ratio} shows the C/O diagnostic in disk atmospheres simulated with the IRIS grid, where models are coded in marker size by stellar mass. Our molecular line ratios are derived from integrated line fluxes of CO$_2$ and H$_2$O for C/H, and of H$_2$O and C$_2$H$_2$ for O/H, using the wavelength regions listed in Table~\ref{table: lines and rms}. For water, we use the forest of lines clustered around $\sim$24~$\mu$m, as this component provides a clearer visual distinction between the different model sets. 

In practice, high (low) H$_2$O/C$_2$H$_2$ line flux ratios indicate a high (low) underlying O abundance, whereas high (low) CO$_2$/H$_2$O ratios indicate a high (low) underlying C abundance. Unlike \citet{Arabhavi_2026}, who varied the elemental C/H and O/H abundances in their models, hence considering a range of sub-solar to super-solar C/O disk ratios, we remark that our models only consider a fixed bulk solar C/O ratio. Nevertheless, the different model sets in the IRIS grid populate distinct regions of the diagnostic plot in Fig.~\ref{figure: C/O ratio}, because the local gas-phase C/O ratio changes accordingly to the different irradiation scenarios, responsible for redistributing C- and O-bearing species in the gas-phase within the disk atmosphere.

In particular, our models predict a range of O/H and C/H ratios clustered around O/H $\sim$0.4--10 and C/H $\sim$1--6, without dramatic dependence on the FUV input from the external environment (see square and star markers in Fig. \ref{figure: C/O ratio}). The scatter among data points seems dominated by the irradiation properties of the central star, and by the difference in thermochemical structure between T\,Tauri and Herbig Ae/Be disks. Overall, our models predict high, carbon-rich C/O atmospheric ratios ($\sim$1--10) for stellar mass regimes. 

Because externally irradiated and stellar-only irradiated disk atmospheres in the IRIS grid exhibit a similar range of carbon-rich C/O ratios with respect to the solar standard, it implies that external external FUV irradiation does not produce a distinct chemical signature. The dependency of our grid on the input elemental abundances of carbon and oxygen plus the effect of gas and dust substructures on the C/O ratio value need future exploration.

\subsection{Model limitations}\label{discussion: model limitations}
Ideally, one would be interested in creating the most realistic disk model possible. However, several physical processes are not included in the present model grid. 

Considering first the absence of a dedicated photoevaporative external wind structure in our models, it is possible to couple ProDiMo with hydrodynamical wind simulations \citep[][]{Rab_2022, Portilla-Revelo_2025}, but this would require running a computationally expensive dedicated hydrodynamical model for each disk structure and subsequently using it as input to ProDiMo. Alternatively, a self-similar wind solution could be adopted, as demonstrated for thermal disk winds by \citet{Sellek_2021}.
Simplified prescription have also been combined with radiative-transfer calculations to model atomic and molecular emission (e.g., H$_2$) observed with JWST from internal disk winds \citep{Sellek_2024, Nakatani_2026}. Similarly for externally driven winds, the recent DALI thermochemical wind modeling by \citet{Keyte_2025, Keyte_2026} show that for $\geq10^{5}$ $G_0$ the synthetic flux of key molecular tracers like CO and CH$_4$ can be enhanced in the outer disk, while transition-dependent changes of H$_2$O line emission are predicted in the upper disk layers due to the temperature enhancement driven by thermal re-emission from the wind. To conclude, future ProDiMo simulations of externally irradiated protoplanetary disks, potentially including the IRIS grid itself, should also treat externally driven photoevaporative winds in order to make more realistic predictions on the rovibrational volatile emission from a wide range of disk radii and layers.

Regarding the assumption of a smooth disk structure, ProDiMo allows the inclusion of multiple gas cavities or dust pressure bumps (see also our zone-model approach in Sect. \ref{subsection: Ext. UV with trunc.}), which has been successfully applied in previous studies \citep[e.g.,][]{Portilla-Revelo_2023, Kanwar_2026}. The main limitation for applying such models to the present study is instead the lack of observational constraints on the location and properties of substructures in protoplanetary disks, isolated and non, and especially at kpc-distance scales. We therefore restrict the current models to smooth disks without substructures, providing a controlled framework to isolate the effects of internal stellar and external irradiation from nearby OB stars. This also lays the basics for a future model grid incorporating a parameterization of disk substructures.

We also stress that our modeling approach of external FUV irradiation does not take into account the expected effects on the dust component. In particular, the dust population should behave differently depending on the grain size, where the small grains get entrained in the photoevaporative wind launched at the outer disk surface, while the inward drift of the larger grains speeds up \citep{Facchini_2016, Sellek_2020, Qiao_2023, Garate_2024, Paine_2025}. Related to this, ProDiMo currently does not allow to take into account the effects of the transport of mm--cm sized dust pebbles \citep{Mah_2023, Mah_2024, Lienert_2024, Lienert_2025}, or the destruction of solid carbonaceous grains \citep{Houge_2025}, both of which have been shown to significantly alter the C/O ratio in the inner disk.

Finally, we note that the present models do not include X-ray radiative transfer through the disk, and therefore neglect the effects of X-ray scattering and absorption by dust and gas on the local radiation field. This represents a potential limitation, as X-ray penetration can modify the thermal and chemical structure of the disk \citep{rab_2018}. Future work should quantify the impact of including X-ray radiative transfer self-consistently on the predicted molecular abundances and line emission, as well as the effects of time-dependent X-ray irradiation.

A large control sample of irradiated T\,Tauri, observed at JWST sensitivity, is needed to further test ProDiMo modeling predictions on rovibrational line fluxes in presence of external FUV irradiation. The very recent JWST/MIRI investigation of a small sample of externally irradiated (10$^3$--10$^5$ $G_0$) T\,Tauri by the 42 Ori B star in NGC 1977 \citep{Booth_2026} already reveals relatively bright and extended H\,\textsc{i} and H$_2$ emission. In comparison with our modeling predictions, the observed flux densities of H\,\textsc{i} with $\Delta n$ = 6--5, 8--6, 7--6 toward KCFF\,\#2 ($M_{\star}$\,$\approx$\,0.4 $M_{\odot}$) and KCFF\,\#6 ($M_{\star}$\,$\approx$\,0.5 $M_{\odot}$) do not allow to distinguish between our baseline or externally FUV irradiated 0.3--0.5 $M_{\odot}$ models. Instead, the observed flux densities of the H$_2$ S(5) and S(4) lines toward these disks agree with the predicted absolute fluxes by FLiTs from the \texttt{Trunc.\,Ext.\,UV} model set, indicating promising diagnostics of disk truncation to be further explored. Taking also into account the near-infrared window observable with the JWST/NIRSpec instrument, further observations will come from other ongoing JWST surveys in NGC 1977 \citep{Romero_2024, Boyden_2024}, as well as the ONC \citep{Rogers_2024, Ballering_2025_proposal, Schroetter_2025_proposal}, and the further out Trumpler 14 cluster \citep{Kuhn_2024}.

In a separate paper, we validate the IRIS grid against already available JWST/MIRI MRS spectra of a sample of 33 T\,Tauri and Herbig Ae/Be disks observed toward both nearby (Taurus, Lupus, Chameleon, Upper-Centaurus Lupus) and clustered environments (NGC 6357). We will demonstrate that despite the above caveats, our grid can correctly predict the trend of increasing line flux density of the most common atomic and molecular mid-infrared gas tracers as a function of stellar mass at the observed sensitivity of JWST. We will also show that our externally irradiated models corroborate the absence of the spectral signature of external FUV irradiation assuming that disks in clustered environments are truncated (Frediani et al., in prep.).

\section{Conclusions}\label{section:Conclusions}
 A first grid of thermochemical 2D disk models of the inner 10 au of protoplanetary disks surrounding young ($\sim$1 Myr) \textit{FGKM}-type pre-main sequence stars of 0.3--4.0 $M_{\odot}$ has been implemented, including variations in the internal (UV and X-ray) and external (FUV) irradiation of the disk. This allows for a parametric investigation of irradiation effects on the innermost physical structure of T\,Tauri and Herbig Ae/Be disks, and their atomic and molecular gas reservoir emitting at infrared wavelengths, inheritable by atmospheres of forming planets.   

Our key findings can be summarized as follows:
\begin{enumerate}

    \item Using stellar and disk properties inferred from the known disk population (i.e., $M_{\star}$, $T_{\rm eff}$, $R_{\star}$, $L_{\rm bol}$, $L_{\rm acc}$, $\dot{M}_{\rm acc}$, $L_{\rm X}$), and assuming a smooth disk structure (no substructures or winds), we consistently predict a linear trend of increasing JWST line luminosity with stellar mass for several key atomic and molecular species  (H\,\textsc{i}, H$_2$, CO, H$_2$O, OH, CO$_2$, C$_2$H$_2$, HCN, and CH$_3^+$). The predicted strengthening of molecular emission toward higher stellar masses highlights a potential tension with \textit{Spitzer} and archival observations of IMTT and Herbig stars, motivating further investigation of how disk structure and irradiation in intermediate-mass star systems affect this trend.
    
    \item We confirm and generalize the result of \citet{Portilla-Revelo_2025} that external FUV irradiation (for 10$^4$ $G_0$) enhances CH$_3^+$ and H$_2$ emission in full disks. In contrast, compact, gas-depleted disks truncated by FUV photoevaporation closely resemble non-irradiated disks, with only small differences in the emission of H\,\textsc{i} (at $\sim$7.5 $\mu$m), H$_2$ ($\sim$6.9--17 $\mu$m), and H$_2$O ($\sim$6--7 $\mu$m). Within this model framework, the absence of the predicted strong CH$_3^+$ and H$_2$ emission in a known irradiated disk supports the interpretation that substantial FUV-induced truncation has already occurred.
       
    \item Mid-IR molecular line ratios (H$_2$O, C$_2$H$_2$, CO$_2$) can trace the fraction of C/O ratio enclosed in the upper warm layers of disk atmospheres. For a fixed total C/O ratio of 0.457 (solar value), accounting for both gas and solids, the IRIS models predict carbon-rich atmospheres (C/O\,$\sim$\,1--10) around both T\,Tauri and Herbig Ae/Be stars. This variation is driven by stellar properties and internal irradiation (but not external FUV fields), which modify the inner disk thermochemical structure and the distribution of C- and O-bearing species emitting in the gas-phase within disk atmospheres.
  
\end{enumerate}
This work highlights the capabilities of thermochemical modeling in characterizing and predicting protoplanetary disk properties, also when anchored to empirical trends observed across the known disk population. The approach here presented also provides a framework for building large, parametrized grids of ProDiMo models in the future, enabling the prediction and validation of observed trends at infrared wavelengths across the disk population rather than on a disk-by-disk basis with JWST, and easily expandable for ELT's first light.

\begin{acknowledgements}
JF, ABi, and ABr acknowledge support from the Swedish National Space Agency (2022-00154).
M.C.R-T acknowledges support by the German Aerospace
Center (DLR) and the Federal Ministry for Economic Affairs and Energy (BMWi) through program 50OR2314 “Physics
and Chemistry of Planet-forming disks in extreme environments.” CH.R
acknowledges the support of the Deutsche Forschungsgemeinschaft (DFG, German Research Foundation) Research Unit
“Transition discs”—325594231. CH.R is grateful for support from the Max Planck
Society. AJW has been supported by the Royal Society through a University Research Fellowship grant number URF\textbackslash R1\textbackslash 241791. 
TP acknowledges support from the Excellence Cluster ORIGINS which is funded by the Deutsche Forschungsgemeinschaft (DFG, German Research Foundation) under Germany’s Excellence Strategy - EXC-2094 - 390783311.
VR acknowledges the support of the European Union’s Horizon 2020 research and innovation program and the European Research Council via the ERC Synergy Grant ``ECOGAL'' (project ID 855130). 

\end{acknowledgements}
\bibliographystyle{bibtex/aa_url}
\bibliography{bibtex/references}
\begin{appendix}
\section{2D disk structure of the IRIS grid}\label{appendix:properties prodimo grids}
\begin{figure*}[!htbp]
    \centering
    \includegraphics[width=0.9\linewidth]{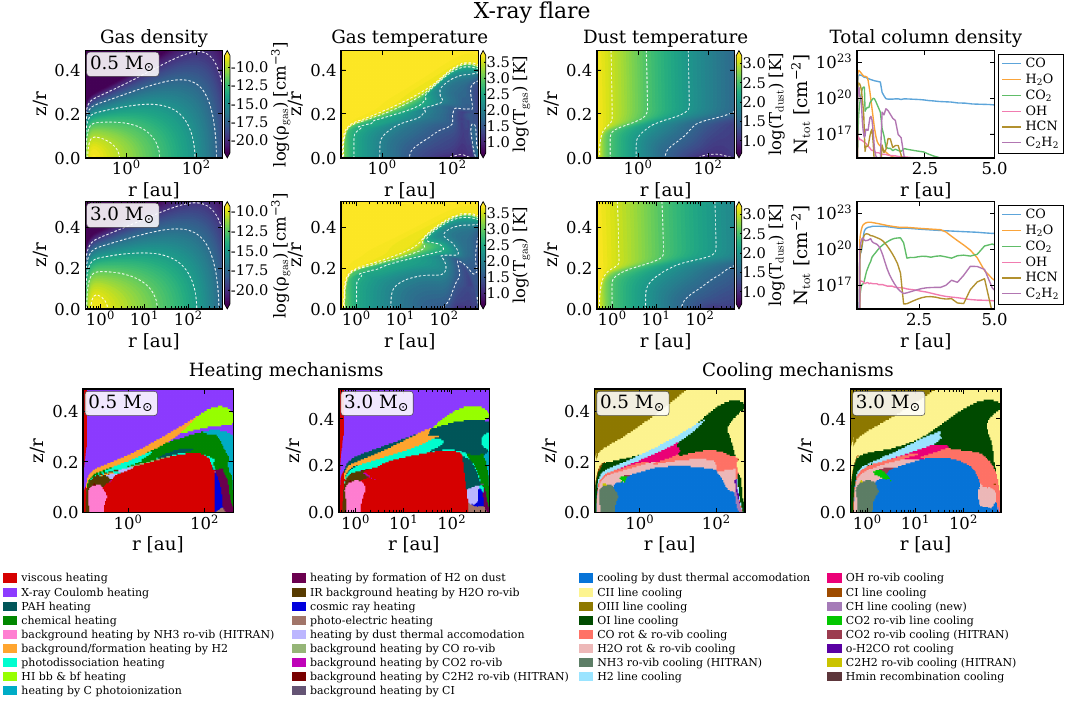}
    \caption{Same as Fig. \ref{figure: baseline grid properties} but for the \texttt{X-ray\,Flare} set of models.}
    \label{figure: xrayFlare grid properties}
\end{figure*}

\begin{figure*}[!htbp]
    \centering
    \includegraphics[width=0.9\linewidth]{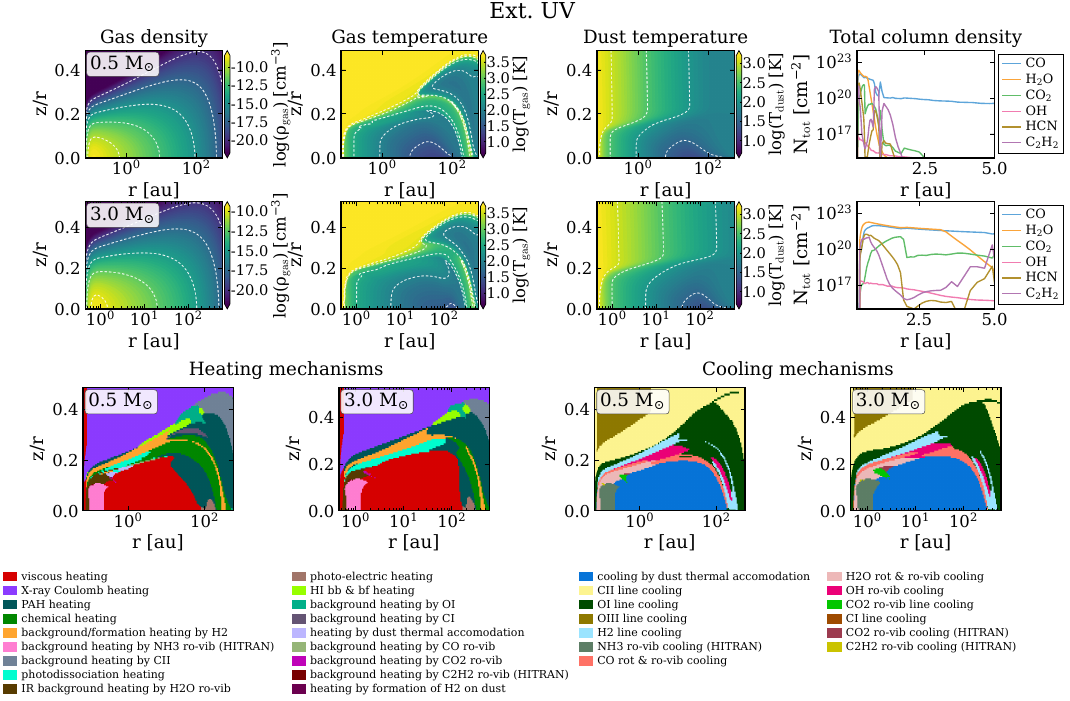}
    \caption{Same as Fig. \ref{figure: baseline grid properties} but for the \texttt{Ext.\,UV} set of models.}
    \label{figure: ExtUV grid properties}
\end{figure*}

\begin{figure*}[!htbp]
    \centering
    \includegraphics[width=0.9\linewidth]{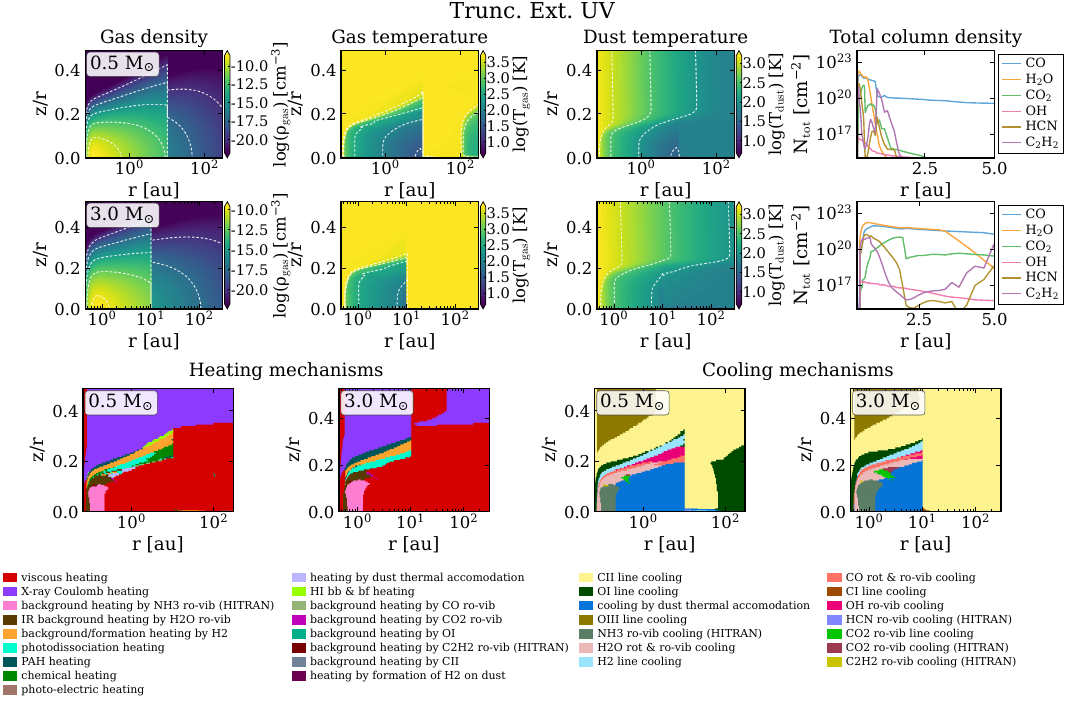}
    \caption{Same as Fig. \ref{figure: baseline grid properties} but for the \texttt{Trunc.\,Ext.\,UV} set of models.}
    \label{figure: Truncated grid properties}
\end{figure*}

Figures \ref{figure: xrayFlare grid properties} to \ref{figure: Truncated grid properties} show the output ProDiMo 2D structure for the \texttt{X-ray\,Flare}, \texttt{Ext.\,UV}, and \texttt{Trunc.\,Ext.\,UV} sets of models, exemplified for a low-mass stellar model ($M_{\star}$ = 0.5 $M_{\odot}$), and for an intermediate-mass stellar model ($M_{\star}$ = 3.0 $M_{\odot}$) in each set. The correspondent figure for the \texttt{Baseline} set is reported in Fig. \ref{figure: baseline grid properties} in the main text. Each figure shows from left to right the 2D gas density profile ($\rho_{\rm gas}$), gas ($T_{\rm gas}$) and dust ($T_{\rm dust}$) 2D temperature profiles, and the total column density ($N_{\rm tot}$) of some of the species analyzed in this work as a function of the radial distance from the star (first and second row), and the main active radiative heating and cooling mechanisms in the disk structure (third row).

\FloatBarrier
\section{Water contamination of synthetic mid-IR spectra}\label{appendix: water contamination of synthetic mid-IR spectra}
\begin{figure*}[!htbp]
 \centering
    \includegraphics[width=0.8\linewidth]{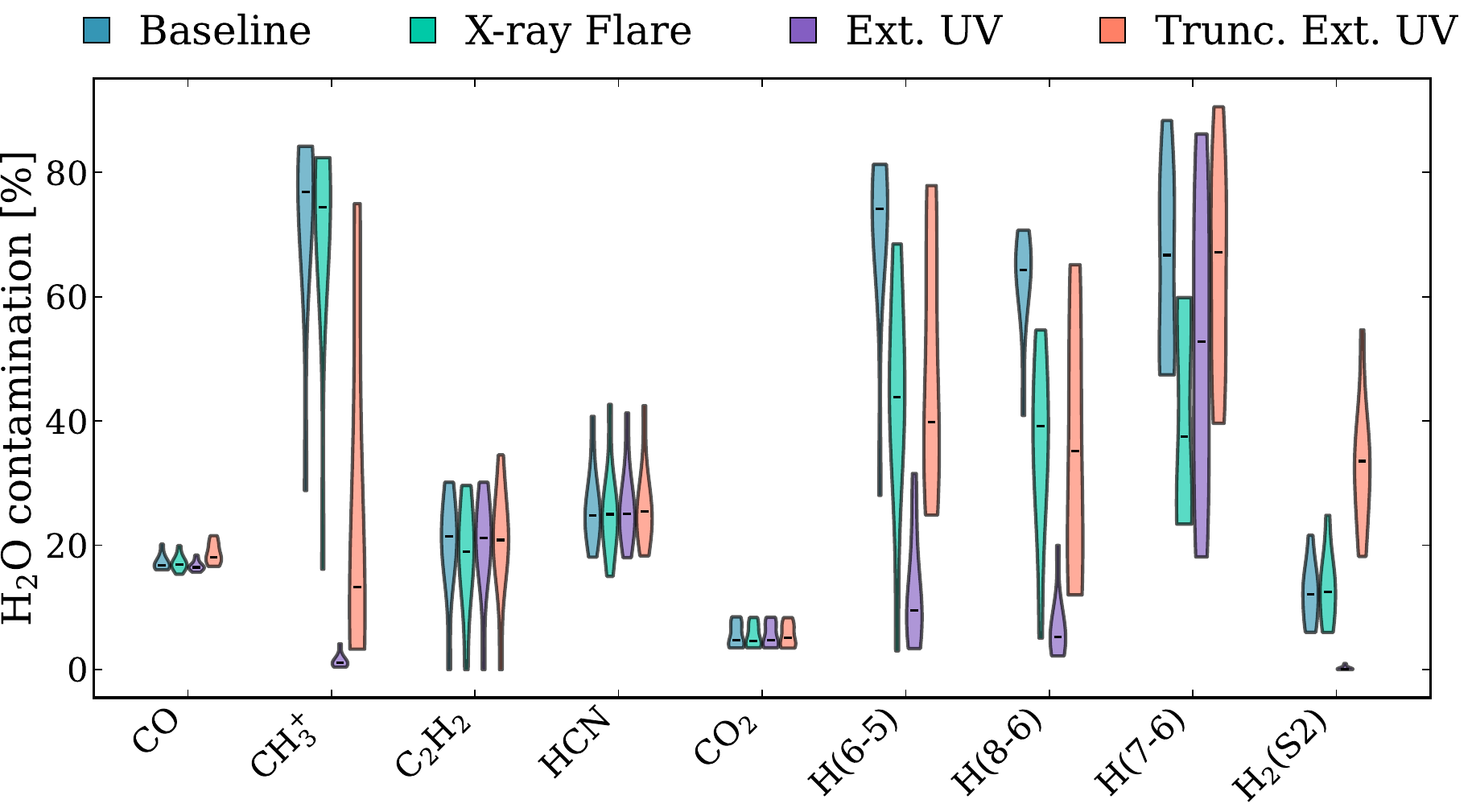}
    \caption{Distribution of H$_2$O contamination across the IRIS grid. The violin plots show the distribution of the fraction of the line flux or molecular band contaminated by H$_2$O emission for each species indicated on the \textit{x}-axis. Each color corresponds to a different set of models (see legend). The central black line indicates the median contamination level, while the width of each violin represents the density distribution of the models within each set.}
    \label{figure: H2O contamination} 
\end{figure*}

To reliably estimate line fluxes of some of the selected species (CO, CH$_3^+$, C$_2$H$_2$, HCN, CO$_2$, H, H$_2$), potentially blended with bright water emission, Fig. \ref{figure: H2O contamination} shows the distributions of the predicted H$_2$O contamination fractions for the different sets of models. The flux water contamination (\textit{y}-axis in figure) is estimated as described in Sect. \ref{section: line flux estimation}. 

\FloatBarrier
\section{FLiTs continuum-subtracted spectra}\label{appendix: FLiTs continuum-subtracted spectra}
\begin{figure*}[!htbp]
 \centering
    \includegraphics[width=\textwidth]{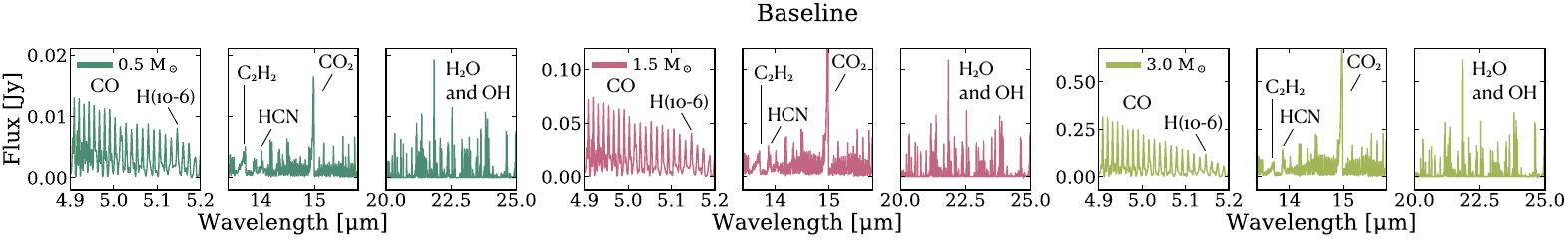}
    \caption{Continuum-subtracted spectral windows of the FLiTs synthetic mid-IR spectra from the \texttt{Baseline} set of models (in Fig. \ref{figure: prodimo spectra baseline}) for increasing stellar host mass (see legend). Relevant atomic and molecular features visible in the spectra are marked in each inset. The \textit{y}-axis scale is fixed for all insets for visualization purposes.}
    \label{figure: baseline spectra zoom}  
\end{figure*}

\begin{figure*}[!htbp]
 \centering
    \includegraphics[width=\textwidth]{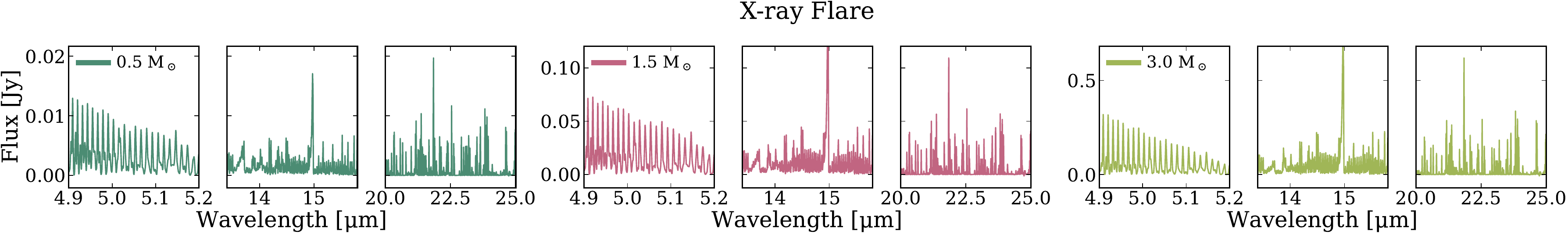}
    \caption{Same as Fig. \ref{figure: baseline spectra zoom} but for the \texttt{X-ray\,Flare} set of models.}
    \label{figure: xrayflare spectra zoom}  
\end{figure*}

\begin{figure*}[!htbp]
 \centering
    \includegraphics[width=\textwidth]{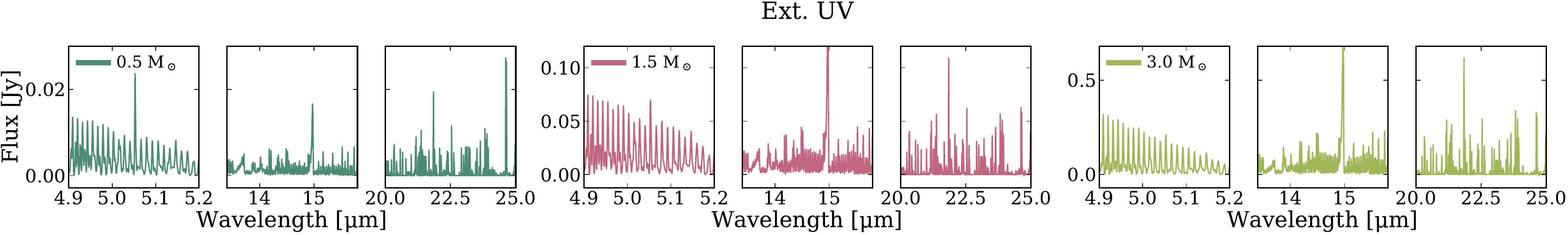}
    \caption{Same as Fig. \ref{figure: baseline spectra zoom} but for the \texttt{Ext.\,UV} set of models.}
    \label{figure: extUV spectra zoom}  
\end{figure*}

\begin{figure*}[!htbp]
 \centering
    \includegraphics[width=\textwidth]{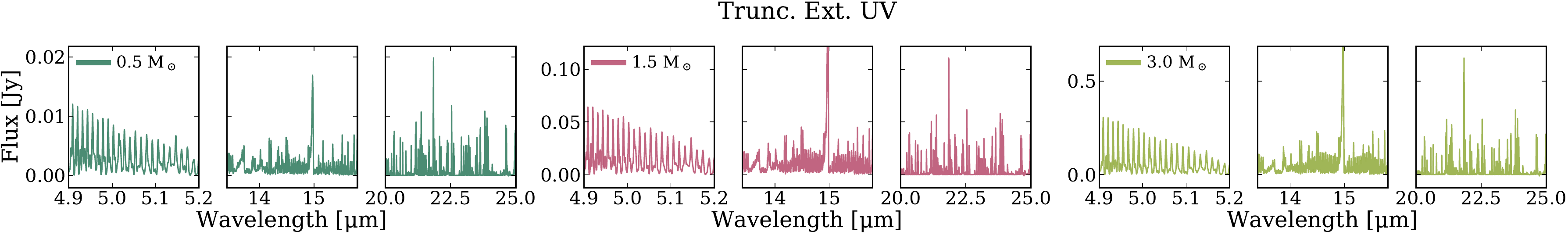}
    \caption{Same as Fig. \ref{figure: baseline spectra zoom} but for the \texttt{Trunc.\,Ext.\,UV} set of models.}
    \label{figure: truncated spectra zoom}  
\end{figure*}

Figures \ref{figure: baseline spectra zoom} to \ref{figure: truncated spectra zoom} show relevant continuum-subtracted spectral windows of the FLiTs spectra in Fig. \ref{figure: prodimo spectra baseline}--\ref{figure: prodimo spectra truncated} produced with the IRIS grid for three representative stellar mass models (0.5, 1.5, 3.0 $M_{\odot}$) in each set.

\FloatBarrier
\end{appendix}
\end{document}